\UseRawInputEncoding
\documentclass{article}
\usepackage[a4paper, total={6in, 10in}]{geometry}
\usepackage[utf8]{inputenc}
\usepackage{xcolor}
\usepackage{graphicx}
\usepackage{lineno}
\usepackage{amsmath}
\usepackage[numbers]{natbib}
\usepackage{ulem}
\usepackage[colorlinks=true,linkcolor=black, citecolor=blue, urlcolor=blue]{hyperref}
\usepackage{subfig}
\usepackage{multirow,multicol}
\usepackage{bm}
\usepackage{soul}
\usepackage{authblk}

\setcitestyle{square}

\title{Prospects for geoneutrino detection with JUNO}

\author[44]{Thomas Adam}
\author[65]{Shakeel Ahmad}
\author[65]{Rizwan Ahmed}
\author[20]{Fengpeng An}
\author[44]{Jo\~{a}o Pedro Athayde Marcondes de Andr\'{e}}
\author[74]{Costas Andreopoulos}
\author[55]{Giuseppe Andronico}
\author[66]{Nikolay Anfimov}
\author[59]{Vito Antonelli}
\author[66]{Tatiana Antoshkina}

% ...
\affil[20]{Sun Yat-Sen University, Guangzhou, China}
% ...
\affil[44]{IPHC, Universit\'{e} de Strasbourg, CNRS/IN2P3, F-67037 Strasbourg, France}
% ...
\affil[55]{INFN Catania and Dipartimento di Fisica e Astronomia dell Universit\`{a} di Catania, Catania, Italy}
% ...
\affil[59]{INFN Sezione di Milano and Dipartimento di Fisica dell Universit\`{a} di Milano, Milano, Italy}
% ...
\affil[65]{Pakistan Institute of Nuclear Science and Technology, Islamabad, Pakistan}
\affil[66]{Joint Institute for Nuclear Research, Dubna, Russia}
% ...
\affil[74]{The University of Liverpool, Department of Physics, Oliver Lodge Laboratory, Oxford Str., Liverpool L69 7ZE, United Kingdom}
% ...

\date{}

\begin{document}
\maketitle

{\small \textbf{Abstract:} Geoneutrinos---antineutrinos emitted during the decay of long-lived radioactive elements inside the Earth---serve as a unique tool for studying our planet's composition and heat budget. The Jiangmen Underground Neutrino Observatory (JUNO) experiment in China, which has recently completed construction, is expected to collect a sample comparable in size to the entire existing world geoneutrino dataset in less than a year.
This paper presents an updated estimation of JUNO's sensitivity to geoneutrinos using the best knowledge available to date about the experimental site, the surrounding nuclear reactors, the detector response uncertainties, and the knowledge of the reactor antineutrino flux.  
 To facilitate comparison with present and future geological models, our results cover a wide range of predicted signal strengths.
Despite the significant background from reactor antineutrinos, the primary channel JUNO will use to determine the neutrino mass ordering, the experiment will measure the total geoneutrino flux with a precision comparable to that of existing experiments within its first few years, ultimately achieving a world-leading precision of about 8\% over ten years.
JUNO's large statistics will also allow separation of the Uranium-238 and Thorium-232 contributions with unprecedented precision, providing crucial constraints on models of Earth’s formation and composition. Observation of the mantle signal above the lithospheric flux will be possible but challenging. For models with the highest predicted mantle concentrations of heat-producing elements, a $\sim$3$\sigma$ detection over six years requires knowledge of the lithospheric flux to within 15\%. Together with complementary measurements from other locations, JUNO’s geoneutrino results will offer cutting-edge, high-precision insights into the Earth’s interior, of fundamental importance to both the geoscience and neutrino physics communities.
}

\tableofcontents

\section{Introduction}\label{sec:intro}

Geoneutrinos are electron antineutrinos ($\bar{\nu}_e$) originating from the radioactive decays of isotopes with half-lives comparable to or longer than Earth's age. These isotopes---$^{232}$Th, $^{238}$U, and $^{40}$K---are naturally present inside the Earth and contribute to a total surface geoneutrino flux of nearly $10^7$\,cm$^{-2}$\,s$^{-1}$~\cite{Fiorentini:2007te,MCDON2020}. Geoneutrinos represent a unique tool for tracing the total abundance, distribution and relative proportion of these so-called {\it Heat-Producing Elements} (HPEs) in the deep Earth. This knowledge provides insights into the Earth's radiogenic heat, a key parameter for geoscience, as well as into the dynamics of mantle convection, mantle homogeneity, and Earth's formation processes. Particularly intriguing is the question of the poorly known radiogenic heat from the Earth's mantle, as HPEs are primarily concentrated in the much better-known Earth's crust, especially in the thick and compositionally complex continental crust~\cite{MCDON2020}. No significant fractions of HPEs are typically expected to be present in the Earth's metallic core due to their chemical affinity for silicates rather than metals.

Geoneutrinos are detected via the {\it Inverse Beta Decay} (IBD) reaction  
$\bar{\nu}_e + p \rightarrow e^+ + n$, a charged-current interaction that is sensitive exclusively to electron antineutrinos. Similar interactions are available to other antineutrino flavors, but reactor and geoneutrinos do not have sufficient energy to produce $\mu^+$ or $\tau^+$ leptons. Organic liquid scintillators, such as the one used in JUNO, provide abundant free target protons for geoneutrino detection. 
The IBD interaction provides a powerful tool to suppress most backgrounds thanks to the strong space and time correlation between the prompt and delayed signals. However, reactor antineutrinos are detected by the very same process and thus represent an irreducible background in geoneutrino measurements. 

The Borexino~\cite{BOREX20} and KamLAND~\cite{Abe2022Geoneutrino} experiments observed geoneutrinos with high statistical significance, detecting 53 and 183 events respectively, and achieving relative precisions on the measured rate in the range of 15–18\%. The observed signals from both experiments---at widely separated locations, with Borexino in Italy and KamLAND in Japan---are consistent with geological predictions, but not precise enough to fully discriminate between different models. {The SNO+ experiment in Canada reported the first detection of geoneutrinos in the Americas~\cite{SNO25} and later improved this measurement, achieving a 26\% precision on the geoneutrino rate~\cite{SNO25-2}.}

The JUNO experiment~\cite{An2016JUNO,JUNOdet}, located near the southern coast of China, will provide the first opportunity to detect geoneutrino signals on the continental margin. With its 20~kton target mass, JUNO will collect unprecedented geoneutrino statistics, surpassing within about a year the combined total sample detected by all other experiments to date. JUNO is located 52.5\,km from two large nuclear power plants, making it an optimal facility for studying the neutrino mass ordering and making precision measurements of neutrino oscillation parameters~\cite{NMO-paper,JUNO_Precision}. Despite the very significant irreducible background from reactor antineutrinos, JUNO is expected to achieve world-leading sensitivity to geoneutrinos shortly after the start of data taking.

JUNO's sensitivity to geoneutrinos was evaluated  previously in Refs.~\cite{An2016JUNO,Han2016}, but nearly nine years have passed since, during which significant progress has been made in detector design and in the understanding of the expected detector performance, reactor antineutrino flux, and backgrounds. This work presents a re-evaluation of JUNO's potential to measure the total geoneutrino signal, to disentangle the relative contributions from $^{232}$Th and $^{238}$U, and to recognize the mantle geoneutrino signal as an excess above the constrained lithospheric signal. Our sensitivity results span a broad range of signal strengths, enabling comparison with a wide variety of predictions from different geological models. 

The paper is structured as follows. Section~\ref{sec:JUNO} presents an overview of the JUNO experiment and the detector design. Geoneutrinos are introduced in Section~\,\ref{sec:geonus}, while Section~\,\ref{sec:geonusAtJUNO} elaborates on expected signal at JUNO site from lithosphere and mantle, building on various existing geological models. Signal event selection and expected backgrounds considered in the analysis are presented in Section~\ref{sec:SignalBackground}. The analysis procedure is explained in Section~\ref{sec:analysis}, the results of the sensitivity studies are shown in Section~\ref{sec:sensitivity}, and a brief conclusion is offered in Section~\ref{sec:conclusions}, and
details of geoneutrino signal calculations at JUNO are provided in Appendix.

\section{The JUNO Experiment}
\label{sec:JUNO}

\begin{figure}[t]
    \centering
    \includegraphics[width=0.8\linewidth]{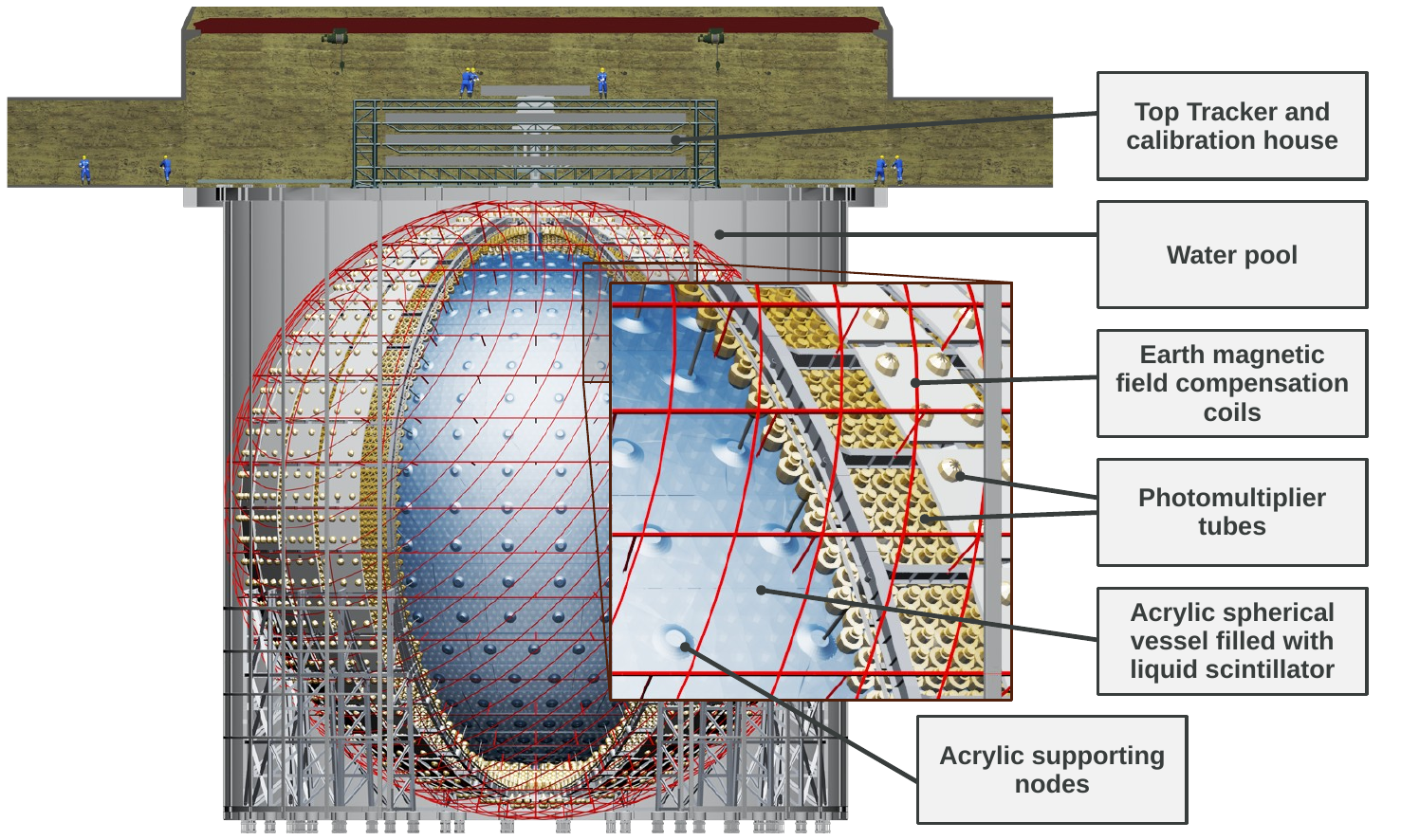}
    \caption{Scheme of the JUNO Detector.}
    \label{fig:detector}
\end{figure}

The Jiangmen Underground Neutrino Observatory (JUNO) is a liquid scintillator detector~\cite{JUNOdet} located at an approximate distance of 52.5\,km from eight nuclear reactors in the Yangjiang and the Taishan nuclear power plants, in an underground laboratory with a vertical overburden of about 650\,m (roughly 1800\,m water equivalent) in the Guangdong Province in Southern China. The experiment's baseline is optimized for the primary goal of determining the neutrino mass ordering (NMO) through the interplay between the fast and slow oscillation patterns of reactor anti-neutrinos~\cite{NMO-paper}, a measurement that will also allow determining three oscillation parameters to sub-percent precision~\cite{JUNO_Precision}. To achieve this, JUNO requires a large target mass and an excellent energy resolution, which offer further opportunities in neutrino and astroparticle physics~\cite{JUNOdet, AtmosphericNeutrinos_JUNO_2021,JUNOB8, JUNOSolar, JUNOSN, DSNB_JUNO_2022,DarkMatter_JUNO_2021,JUNOProton}. The JUNO collaboration will also deploy a satellite detector called the Taishan Antineutrino Observatory (TAO) that will provide a reference reactor antineutrino spectrum for JUNO's oscillation measurements~\cite{TAO}. 

A sketch of the JUNO detector is shown in Fig.~\ref{fig:detector}. A Central Detector (CD)~\cite{JUNOCD} contains the primary target consisting of 20\,kton of a liquid scintillator mixture in a 17.7\,m radius acrylic sphere. This makes JUNO the largest liquid scintillator detector ever built, in comparison with the 280\,tons, 780\,tons, and 1\,kton active masses of Borexino~\cite{Bxdet}, SNO+~\cite{SNO+}, and KamLAND~\cite{KamLAND:2002uet}, respectively. The linear alkylbenzene (LAB)-based liquid scintillator mixture was optimized in dedicated studies using a Daya Bay detector~\cite{DBLS}. 
The acrylic vessel (AV) is supported by a spherical stainless steel (SS) structure via 590 connecting bars. Scintillation light is detected by 17,612 20-inch photomultiplier tubes (PMTs)~\cite{JUNOLPMts} and 25,600 3-inch PMTs facing the acrylic sphere that are mounted on the SS structure, providing a total photocathode coverage of 78\%. These unique features lead to an energy resolution of $\approx 3\%/ \sqrt{E(\mathrm{MeV})}$~\cite{JUNOResol}, which is unprecedented for a detector of this type. 

The CD is installed inside a cylindrical water pool (WP) 43.5\,m in diameter and 44.0\,m in height that is filled with 41\,kton of ultra-pure water. The WP shields the CD against external radiation and acts as a Cherenkov veto detector for the residual cosmic-ray muon flux of 0.004\,m$^{-2}$s$^{-1}$. The veto system contains 35\,kton of water, while an additional 6\,kton of water are used in the CD region between the SS and the AV. The WP veto is equipped with 2,400 20-inch PMTs installed on the outer surface of the SS structure. On top of the WP, a Top Tracker (TT)~\cite{JUNOTT} is placed to precisely measure the tracks of a sub-sample of crossing muons.

A comprehensive multi-positional calibration system relying on various artificial and natural radioactive sources, as well as a pulsed laser, has been developed to calibrate the detector's spatial non-uniformity and energy scale non-linearity to sub-percent precision~\cite{JUNOcalib}. 

Detector construction was completed in late 2024, and liquid scintillator filling of the central target was completed on August 22, 2025.

\section{Geoneutrinos}
\label{sec:geonus}

\begin{figure}[t]
    \centering
    \includegraphics[width=0.7\linewidth]{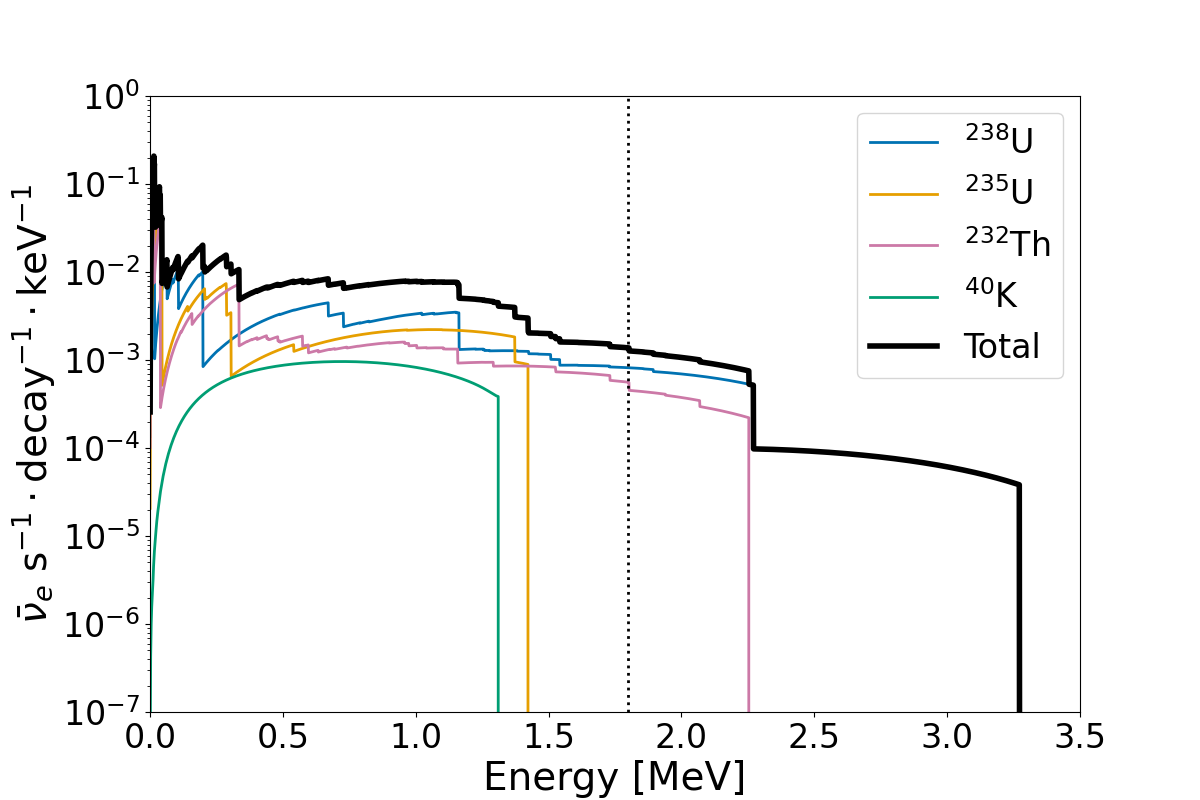}
    \caption{Geoneutrino energy spectra~\cite{Enomoto} from the decays of the $^{238}$U, $^{235}$U, and $^{232}$Th chains and of $^{40}$K. The total spectrum integrals are 6 for $^{238}$U, 4 for $^{235}$U and $^{232}$Th, and 0.89 for $^{40}$K. The vertical dotted line indicates the kinematic threshold of the IBD interaction.}
    \label{fig:GeoNuDecay}
\end{figure}

Geoneutrinos are emitted by HPEs that have persisted since the planet's formation from primordial material, primarily
 $^{232}$Th ($T_{1/2}$ = 1.41(1)\,$\cdot$\,10$^{10}$\,years), $^{238}$U ($T_{1/2}$ = 4.4683(96)\,$\cdot$\,10$^{9}$\,years), $^{235}$U ($T_{1/2}$ = 7.0348(20)\,$\cdot$\,10$^{8}$\,years), and $^{40}$K ($T_{1/2}$ = 1.262(2)\,$\cdot$\,10$^{9}$\,years)~\cite{MCDON2020}. Natural thorium is fully composed of $^{232}$Th, while natural uranium, largely dominated by $^{238}$U, contains about 0.7204\% of $^{235}$U. Natural potassium contains only about 0.0117\% of $^{40}$K. Other HPEs exist, but their contributions are subdominant. In each decay, the emitted {\it radiogenic heat} is in a well-known ratio to the number of emitted geoneutrinos:
\begin{eqnarray}
&^{238}\mathrm{U} \rightarrow \, ^{206}\mathrm{Pb} + 8\alpha + 6 e^{-} + 6 \bar{\nu}_e,\, \mathrm{Q} = 51.7\,\mathrm{MeV}\label{Eq:geo1}\\
&^{235}\mathrm{U} \rightarrow \, ^{207}\mathrm{Pb} + 7\alpha + 4 e^{-} + 4 \bar{\nu}_e,\, \mathrm{Q} = 46.4\,\mathrm{MeV} \label{Eq:geo2}\\
&^{232}\mathrm{Th} \rightarrow \, ^{208}\mathrm{Pb} + 6\alpha + 4 e^{-} + 4 \bar{\nu}_e,\, \mathrm{Q} = 42.6\,\mathrm{MeV} \label{Eq:geo3} \\
&^{40}\mathrm{K} \rightarrow \, ^{40}\mathrm{Ca}  +  e^{-} +  \bar{\nu}_e,\, \mathrm{Q} = 1.31\,\mathrm{MeV}~\mathrm{(89.3\%)} \label{Eq:geo4}\\
&^{40}\mathrm{K} + e^{-} \rightarrow \, ^{40}\mathrm{Ar}  +  \nu _{e},\, \mathrm{Q} = 1.504\,\mathrm{MeV}~\mathrm{(10.7\%)} \label{Eq:geo5}
\end{eqnarray}

The energies listed above represent total decay energies, of which geoneutrinos carry approximately 5\% in the form of kinetic energy~\cite{Fiorentini:2007te}.
The corresponding energy spectra~\cite{Enomoto}, shown in Fig.~\ref{fig:GeoNuDecay}, all lie below 3.27\,MeV. We note that the branch of the $^{238}$U spectrum extending up to about 4.4\,MeV is roughly three orders of magnitude lower and can therefore be considered negligible for the present study.

As mentioned in Section~\ref{sec:intro}, the IBD reaction yields a positron and a neutron. The positron first deposits its kinetic energy before annihilating into two 511\,keV $\gamma$-rays, generating a prompt signal with visible energy $E_{p}$ that is directly correlated with the antineutrino energy $E_{\bar{\nu}_e}$: 
\begin{equation}
E_{p} \sim E_{\bar{\nu}_e}- 0.78\,\, \mathrm {MeV}.
\label{eq:Epro}
\end{equation}
The capture of the thermalized neutron
occurs primarily on a free proton or with $\sim$1\% probability on $^{12}$C.
It is followed by a $\gamma$-ray emission (2.22\,MeV or 4.9\,MeV for capture on H or $^{12}$C, respectively) that constitutes the delayed signal. The tight temporal and spatial coincidence between the prompt and delayed signals provides a clear signature for $\bar{\nu}_e$ events that allows to effectively suppress backgrounds.

The IBD cross-section is precisely known, with an uncertainty of about 0.4\%~\cite{strumia2003precise}. The reaction is sensitive only to the electron neutrino flavor and has a kinematic threshold of 1.806~MeV. As shown in Fig.~\ref{fig:GeoNuDecay}, $^{40}$K and $^{235}$U geoneutrinos fall below the IBD threshold. Only 0.38 and 0.15 antineutrinos per decay of $^{238}$U and $^{232}$Th, respectively, exceed this threshold, with maximal energies of 3.27 MeV and 2.25 MeV. 

The geoneutrino signal is often expressed in {\it Terrestrial Neutrino Units} (TNU), which simplifies the conversion between geoneutrino flux and IBD event rates.  One TNU corresponds to one IBD event per year in a detector with 100\% efficiency and 10$^{32}$ free target protons (approximately 1\,kton of LS).

Estimating the abundance of a given HPE in the bulk Earth relies on modeling. {\it Bulk Silicate Earth} (BSE) models integrate cosmochemical, geochemical, and geophysical data to estimate the composition of the primordial bulk silicate Earth, after the separation of the metallic core. While there are several categories of BSE models (see also Section~\ref{subsec:mantle}), they typically predict a few $10^{16}$\,kg of uranium, $10^{17}$\,kg of thorium, and $10^{20-21}$\,kg of potassium~\cite{BOREX20}.  

Considering the natural abundances of radioactive isotopes and their specific radiogenic heat, the Earth's total radiogenic heat is predicted to range from about 10 to 37\,TW~\cite{Bell21}. All BSE models predict an almost equal contribution from U and Th to the radiogenic heat, while $^{40}$K is expected to contribute about 14--17\% due to its lower elemental specific heat production. The corresponding average surface geoneutrino flux is on the order of $10^7$\,cm$^{-2}$\,s$^{-1}$, which yields a geoneutrino signal $S$ ranging from a few to several tens of TNU, depending on the location.  

Measurement of the geoneutrino signal can provide important information about the Earth's radiogenic heat, HPE distribution and, consequently, the processes involved in Earth's formation. While different BSE models predict HPE abundances that vary by a factor of 2 to 3, all models predict similar proportions of HPEs. This consistency is based on the observed strong correlation of isotopic ratios in the solar photosphere and meteorites~\cite{Bellini2013}. This is particularly important given that present-day detection techniques are not yet capable of measuring $^{40}$K geoneutrinos. {An experimental confirmation of the Th/U ratio consistent with theoretical expectations would validate existing models and increase confidence in extrapolating the Earth’s total radiogenic heat from U, Th, and K based solely on measurements of U and Th geoneutrinos.}

The primordial BSE later differentiated into the present-day crust and mantle. This process redistributed HPEs, which are characterized by high concentrations in the crust. For a more precise estimation of the geoneutrino signal expected at a specific location, several geological factors, as well as the local geology around the experimental site, must be considered. This is broadly discussed in the next section for the case of JUNO in South China.  

\section{Expected Geoneutrino Signal at JUNO}
\label{sec:geonusAtJUNO}

The evaluation of the expected geoneutrino signal requires knowledge of the abundances and distribution of HPEs inside the Earth as the key inputs. Due to their geochemical affinity, HPEs concentrated in the silicate part of the Earth after the formation of the metallic core~\cite{MCDON95,WOOD}. During the differentiation of the silicate Earth, HPEs were concentrated in the continental crust, as their ionic properties favor partitioning into silicate melts and upward transport by magmatism. While the composition of the more accessible, shallow layers of the Earth can be constrained through rock sample analysis and high-resolution geophysical data, the abundances of elements in the mantle are inferred indirectly from geochemical arguments and BSE compositional models.

For detectors situated on the continental crust, such as JUNO, the geoneutrino signal is predominantly influenced by the area within a few hundred kilometers around the experimental site~\cite{RRM}. In this work we present JUNO's sensitivity to geoneutrinos in a model independent way. In order to anchor the range of considered expected signals, we provide an overview of different models that predict the signal from the lithosphere (Section~\ref{subsec:litho})
and mantle (Section~\ref{subsec:mantle}), with further details provided in the Appendix. Figure~\ref{fig:lithos_model} presents a schematic of the Earth's structure highlighting the components important for the geoneutrino signal calculation. 

The oscillation of geoneutrinos during their propagation from their production site to JUNO must, in principle, be considered. 
However, even though studies performed between 2015 and 2020 used different oscillation parameter sets
available at the time, their impact is {\color{blue} small} compared to the much larger uncertainties in geological models and input assumptions.

\subsection{Lithospheric contribution}
\label{subsec:litho}

The lithosphere consists of the Earth’s crust and the underlying Continental Lithospheric Mantle (CLM), which is mechanically coupled to the crust but significantly poorer in HPEs.
For detectors situated on the continental crust, such as JUNO, the geoneutrino signal is predominantly (40-80\%) influenced by the adiajcent lithosphere, primarily from the Local Crust (LOC), the region within a few hundred kilometers of the detector. Detailed geological information from the LOC has a larger impact on the prediction with respect to the Rest Of the Crust (ROC) and CLM.

\begin{figure}[t]
    \centering
    \includegraphics[width=0.8\linewidth]{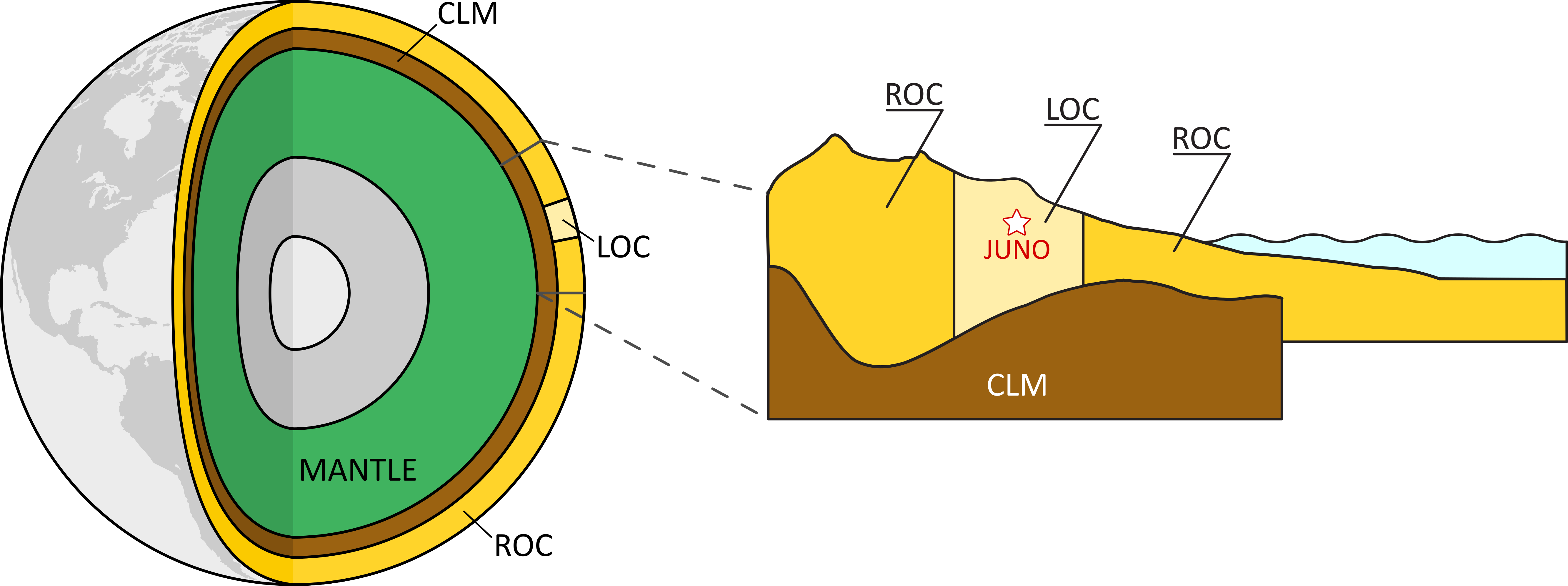}
    \caption{Schematic drawing (not to scale) of the Earth’s structure highlighting the components contributing to the geoneutrino signal expected at JUNO: the mantle, the Continental Lithospheric Mantle (CLM) and the crust. The crust is divided into the Local Crust (LOC), defined as the portion within a few hundred kilometers from the detector, and the Rest of the Crust (ROC).}
    \label{fig:lithos_model}
\end{figure}

For this study, the geoneutrino signals from the LOC, ROC, and CLM reported in previous works were analyzed (Figure~\ref{fig:lithos_model}). All models employ different methodologies and approaches with variations in the size and coordinates of the LOC but remain strongly correlated due to partially shared geophysical and geochemical inputs. In particular, we considered the following models:
\begin{enumerate}
\item {The global Refined Reference Earth Model (RRM)~\cite{RRM} for HPEs' abundances and distribution is exploited for the prediction at JUNO site in Ref.~\cite{Strati15}}.
\item {GIGJ (GOCE Inversion for Geoneutrinos at JUNO)~\cite{GIGJ} is a 3D model exploiting local gravimetric data (GOCE), seismic data, and HPE abundances from RRM. 
\item{Wipperfurth et al. (W20) \cite{W20} considers three different geophysical models CRUST2.0~\cite{CRUST2} (W20 [C2.0]), CRUST1.0~\cite{CRUST1} (W20 [C1.0]), and LITHO1.0~\cite{LITHO} (W20 [L1.0]), while the HPE abundances are derived using a strategy similar to the RRM.
} 
\item{JULOC~\cite{JULOC} incorporates a 3D model of the LOC using as geophysical input a 3D shear wave velocity model inverted from seismic ambient noise tomography. The geochemical inputs are based on U and Th abundances from over 3000 rock samples in the region and from global U and Th databases.}
}
\end{enumerate}

\begin{figure}[t]
    \centering    \includegraphics[width=0.6\linewidth]{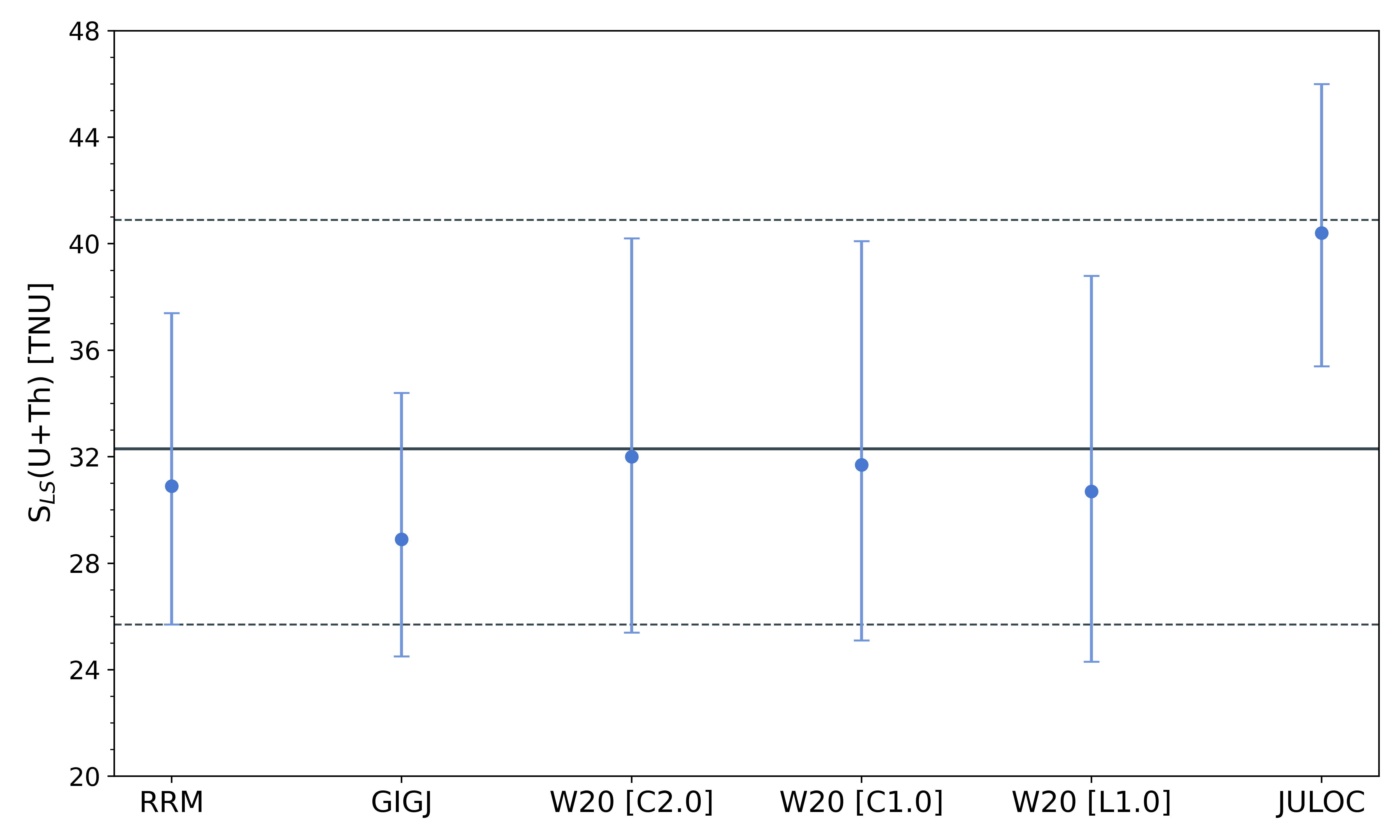}
    \caption{Predicted geoneutrino signals $S$(U+Th) for the six lithospheric models, with error bars representing the corresponding $1\sigma$ uncertainties. The solid horizontal line marks the central value of the combined signal, while the dashed lines indicate its $1\sigma$ bounds.}
    \label{fig:sig_lithos}
\end{figure}

Figure~\ref{fig:sig_lithos} compares the prediction of all six models. Most models report a lithospheric geoneutrino signal $S_\text{LS}(\text{U}+\text{Th})$ with a central value in the range of 28.9\,TNU to 32.0\,TNU, with an average relative uncertainty of about 20\% (Appendix, Table~\ref{tab:signals_ls}). Instead, JULOC stands out with a much larger signal of $S_\text{LS}(\text{U}+\text{Th}) = 40.4\,\text{TNU}$ and a lower relative uncertainty of about 10\%.
The ratios of S(Th)/S(U) remain consistent across all models, with values clustering around 0.30–0.31. This stability reflects the strong correlation between U and Th contributions and highlights the uniformity in the assumed Th/U abundance ratio across the models.

Since all these models  share some common features, it is necessary to account for strong correlations when calculating a single central value and the corresponding uncertainty for the expected lithospheric signal at JUNO. A combined prediction has been derived by a study in which 
 $10^6$ values of geoneutrino signals were generated for each model. The resulting combined prediction for the lithospheric geoneutrino signal at JUNO was $S_{\mathrm{LS}}$(U) = $24.7^{+6.7}_{-5.5}$\,TNU and
$S_{\mathrm{LS}}$(Th) = $7.5^{+2.2}_{-1.8}$\,TNU. The combined prediction was $S_{\mathrm{LS}}$(U + Th) = $32.3^{+8.6}_{-6.6}$\,TNU, with
a signal ratio 
$S_{\mathrm{LS}}$(Th)/$S_{\mathrm{LS}}$(U) of 0.30 (Appendix, Table~\ref{tab:signals_ls}).

\subsection{Mantle contribution}
\label{subsec:mantle}

 Since direct access to the deep mantle is impossible, the BSE models integrate cosmochemical, geochemical, and geophysical data to estimate the HPE distribution. Cosmochemical constraints are grounded on chondritic meteorites, particularly carbonaceous ~\cite{MCDON16} and enstatite chondrites~\cite{JAV10}, considered proxies for the building blocks of Earth. Also, geochemical data from Mid-Ocean Ridge Basalts provide direct insights into the upper mantle’s composition~\cite{KORE08}. Geophysical evidence from seismic wave velocities, surface heat flow and mantle density profiles further constrain mantle composition~\cite{TURCO18}. 

A broad range of BSE compositional models has been proposed, which can be categorized based on their predicted radiogenic heat production (H) into three classes: (1) poor-H models, (2) medium-H models, and (3) rich-H models. Poor-H models, based on enstatite chondrites, predict low U and Th abundances, resulting in a total radiogenic heat $H_\text{BSE}^\text{Poor} = 12.4 \pm 1.9~\text{TW}$. Medium-H models assume relative refractory lithophile element abundances similar to CI chondrites, calibrated with terrestrial samples, and predict $H_\text{BSE}^\text{Medium} = 19.7 \pm 3.1~\text{TW}$. Rich-H models, which include geodynamic models based on mantle convection energetics, predict higher radiogenic heat outputs at $H_\text{BSE}^\text{Rich} = 31.7 \pm 3.4~\text{TW}$ (Table 17 of Ref.~\cite{Bell21}).

Once the total U and Th content is fixed for a given BSE class, and the lithospheric U and Th abundances are determined ($H_\text{LS} = 6.9_{-1.2}^{+1.6}~\text{TW}$),~\cite{RRM}), the residual amounts of U and Th in the mantle are calculated by subtracting the lithospheric contribution from the total BSE estimates (Table\,29 of Ref.~\cite{Bell21}). The mantle radiogenic heat obtained for each BSE class (Table~\ref{tab:mantle_signal_text}) reflects the assumptions regarding the Earth’s primordial composition and an assumed model for the LS. The expected mantle geoneutrino signal is obtained by applying a linear relationship between the radiogenic heat generated by U and Th in the mantle and the corresponding signal. The proportionality coefficient accounts for the spatial distribution of HPEs and is independent of their absolute abundance. For each BSE class, the uncertainty on the signal $S_M(\mathrm{U}+\mathrm{Th})$, as reported in Table~\ref{tab:mantle_signal_text}, reflects both the compositional variability within each model and the uncertainty related to the distribution of HPEs in the mantle. The methodology outlined in the Appendix enables a consistent estimation of the expected signal ranges for the various BSE models. As a consequence, the relative uncertainties vary significantly: the Poor-H model exhibits the highest average relative uncertainty ($\sim$80\%), whereas the Rich-H model shows the lowest ($\sim$30\%) (Table~\ref{tab:mantle_signal_text}).

Given its proportionality to the mass ratio $M_\text{Th}/M_\text{U}$, the signal ratio $S_\text{M}(\text{Th})/S_\text{M}(\text{U})$ varies across the compositional models, with the Poor-H class reporting the lowest ratio (0.17), corresponding to a higher relative U abundance (see Section 2.1 in~\cite{Fiorentini:2007te}). In contrast, the relatively higher Th abundance in the Rich-H class leads to the highest value of $S_\text{M}(\text{Th})/S_\text{M}(\text{U}) = 0.26$. We note that this variability is primarily a consequence of different mobility of uranium and thorium and different enrichment in the crust over geological time, while the bulk $M_\text{Th}/M_\text{U}$ in all BSE models is close to a chondritic value of about 3.9~\cite{Bell21, BOREX20,WRATIO}.

It is worth mentioning that the uncertainty of the lithospheric signal $\left(S_\text{LS} = 32.3_{-6.6}^{+8.6}\,\text{TNU}\right)$ is larger than the central value of the Poor-H mantle signal $\left(S_\text{M} = 2.8_{-1.9}^{+2.4}\,\text{TNU} \right)$ and comparable to that of the Medium-H mantle signal $\left(S_\text{M} = 8.0_{-3.2}^{+4.0}\,\text{TNU} \right)$. This highlights the critical need for refining the local crustal model and reducing the lithospheric uncertainty to enable identification of the mantle signal above the constrained lithospheric contribution, as discussed in Sec.~\ref{sec:ResultsMantle}.

\begin{table}[t]
    \centering
        \renewcommand{\arraystretch}{1.3}
   \begin{tabular}{lccc}
        \hline\hline
        \textbf{} 
          &Poor-H & Medium-H & Rich-H \\
        \hline\hline
        \multicolumn{4}{c}{Mantle radiogenic heat [TW]}\\ \hline
        H$_M$(U+Th+K) & $4.2^{+2.4}_{-2.6}$ & $11.4^{+3.5}_{-3.6}$ & $23.4^{+3.8}_{-3.9}$ \\
        $H_M$(U+Th) & $3.2^{+2.0}_{-2.1}$ & $9.3^{+2.9}_{-2.9}$ & $20.2^{+3.2}_{-3.3}$ \\ \hline \hline
    \multicolumn{4}{c}{Mantle geoneutrino signal [TNU]} \\
        \hline
        $S_M$(U) & $2.4^{+2.1}_{-1.7}$ & $6.5^{+3.2}_{-2.6}$ & $13.7^{+4.5}_{-3.7}$ \\
        $S_M$(Th) & $0.4^{+0.3}_{-0.3}$ & $1.5^{+0.7}_{-0.6}$ & $3.6^{+1.2}_{-1.0}$ \\
        $S_M$(U+Th) & $2.8^{+2.4}_{-1.9}$ & $8.0^{+4.0}_{-3.2}$ & $17.4^{+5.7}_{-4.7}$ \\ \hline
        \multicolumn{4}{c}{Mantle geoneutrino signal ratio} \\ \hline
        S(Th)/S(U) & 0.17 & 0.23 & 0.26 \\
        \hline\hline
    \end{tabular}
    \caption{Mantle radiogenic heat production from U, Th and K ($H_\text{M}(\text{U}+\text{Th}+\text{K})$) and from U and Th alone ($H_\text{M}(\text{U}+\text{Th})$) for the three classes of Bulk Silicate Earth (BSE) model: Poor-H, Medium-H and Rich-H. The mantle geoneutrino signal ($S_\text{M}(\text{U}+\text{Th})$) is derived using the linear relationship discussed in Appendix. The U and Th components are calculated based on the relation between the signal ratio ($S_\text{M}(\text{Th})/S_\text{M}(\text{U})$) and the mass ratio of Th to U ($M_\text{Th}/M_\text{U}$), the latter predicted by BSE.}
    \label{tab:mantle_signal_text}
\end{table}

\subsection{Total Signal}
\label{subsec:TotalSignal}

The total geoneutrino signal at JUNO is determined by combining the lithospheric contribution $S_\text{LS} = 32.3_{-6.6}^{+8.6}$\,TNU, calculated based on estimates from published models, with the mantle signal, evaluated according to the three compositional classes of BSE (Table~\ref{tab:mantle_signal_text}). This approach results in three distinct total geoneutrino signals: one representing to the Poor-H model, one for the Medium-H model, and one for the Rich-H model (Table~\ref{tab:signals_tot}). The corresponding total geoneutrino signal, including uncertainties, spans the range from 28.6\,TNU to 60.3\,TNU. This interval is fully covered in our scan of JUNO’s sensitivity to different geoneutrino signal strengths ranging from 20 to 70 TNU, as presented in Sections~\ref{sec:ResultsUThFixed} and~\ref{sec:ResultsUThFree}.

The uncertainties on the total geoneutrino signals for the Poor-H, Medium-H, and Rich-H models from Table~\ref{tab:signals_tot} were calculated by treating the contributions from the lithosphere and the mantle as linearly independent, taking into account that the errors associated with the lithospheric signal and those linked to the mantle signal are uncorrelated. The asymmetrical uncertainties, reflecting the non-Gaussian nature of the distributions, are propagated using the probability density functions of each component. We note that for each model, the total uncertainty is dominated by the uncertainty associated with the lithospheric contribution, which reflects variability in the crustal model and published estimates. The roughly $\pm$7\,TNU spread of the central values among the three models represents the variability of the mantle signal prediction. 

The individual contributions of U and Th to the total geoneutrino signal have been computed, allowing the calculation of the S(Th)/S(U) ratios. Independently of the considered BSE class, the ratio of the total signal remains constant at 0.29 (Table~\ref{tab:signals_tot}), as it is predominantly controlled by the lithospheric contribution, which strongly influences the thorium-to-uranium signal balance.

\begin{table}[t]
    \centering
    \renewcommand{\arraystretch}{1.3}
    \begin{tabular}{ccccc}
       \hline\hline
       &S(U) & S(Th) & S(U+ Th) & S(Th)/S(U)\\
       &[TNU] & [TNU] &  [TNU] & \\
       \hline\hline
        Poor-H & $28.1^{+8.2}_{-6.2}$ & $8.1^{+2.3}_{-1.8}$ & $36.0^{+9.6}_{-7.4}$ & 0.29\\
        Medium-H & $32.0^{+7.9}_{-6.4}$ & $9.1^{+2.4}_{-1.9}$ & $41.0^{+9.5}_{-7.7}$ & 0.29 \\
        Rich-H & $39.1^{+8.3}_{-6.9}$ & $11.3^{+2.5}_{-2.1}$ & $50.2^{+10.1}_{-8.5}$ & 0.29 \\
        \hline\hline
    \end{tabular}
    \caption{Total geoneutrino signals obtained summing the combined lithospheric signal and the mantle signals for the three BSE models (Poor-H, Medium-H, Rich-H). The uncertainties include contributions from the lithosphere and the mantle, treated as independent. The U and Th components are reported together with their ratio.}
    \label{tab:signals_tot}
\end{table}

\section{Event Selection and Expected Backgrounds}
\label{sec:SignalBackground}

The space and time coincidence signature of the IBD interaction, used to detect geoneutrinos, provides excellent means of background suppression. The selection criteria for IBD candidates are described in Section~\ref{subsec:IBDselection}. Reactor antineutrinos, detected through the same IBD process, constitute an irreducible background in geoneutrino detection, as will be detailed in Section~\ref{subsec:ReaBgr}. Other non-antineutrino backgrounds must be thoroughly evaluated too, as discussed in Section~\ref{subsec:OtherBgs}.

\subsection{IBD Event Selection}
\label{subsec:IBDselection}

Geoneutrinos are detected via the same IBD reaction as reactor antineutrinos and the analysis of both is performed on the prompt energy spectrum from the IBD threshold up to the end point of reactor spectrum (see Section~\ref{sec:analysis}). Consequently, the selection of IBD event candidates is the same for geoneutrino analysis, as in the previous JUNO sensitivity studies regarding reactor antineutrinos~\cite{NMO-paper,JUNO_Precision}. A fiducial volume (FV) cut of 17.2\,m away from the detector center (within a distance of 0.5\,m from the inner
surface of the AV) is applied in order to reduce the external background. The prompt IBD candidate events are restricted to the energy window [0.8, 12.0]\,MeV.
The delayed candidate must have energies in the windows of [1.9, 2.5] MeV or [4.4, 5.5] MeV, corresponding to the capture on Hydrogen and Carbon, respectively. Then, a correlation between a prompt and delayed candidate is searched by requiring a separation of no more than 1.0\,ms in time and 1.5\,m in space. This will greatly reduce the contribution of uncorrelated events. Lastly, a complex veto for cosmic muons and cosmogenic events, in particular neutrons and $^9$Li/$^8$He, is applied. When all cuts are applied in sequence, the overall tagging efficiency for IBD events is 82.2\% including the effect of FV (91.5\%) and muon veto (91.6\%).

\subsection{Reactor Antineutrino Background}
\label{subsec:ReaBgr}

\subsubsection{Nearby reactors within 300\,km}
\label{subsec:localrea}

Reactor antineutrinos from the nearby power plants constitute the largest background for the geoneutrino measurement, with a dominant contribution from the Taishan and Yangjiang power plants. This background also includes antineutrinos from Daya Bay power plant at a distance of 215\,km.
For each core, the antineutrino flux expected at JUNO from the fission of $^{235}$U, $^{238}$U, $^{239}$Pu, and $^{241}$Pu is estimated from the Huber-Mueller model~\cite{Huber2011Erratum, Mueller2011}, weighed with the IBD cross section~\cite{strumia2003precise}, and corrected with the Daya-Bay observation~\cite{DayaBay2017}. Fission fractions of 0.58, 0.07, 0.30, and 0.05 for $^{235}$U, $^{238}$U, $^{239}$Pu, and $^{241}$Pu, respectively, as well as the $Q$ values of the decays of these isotopes, are taken from Ref.~\cite{QValues}. Corrections accounting for non-equilibrium and spent nuclear fuel contributions are applied to the spectra. This procedure, which is the same used in Refs.~\cite{NMO-paper,JUNO_Precision}, yields a reactor antineutrino rate of 43.2 events per day (1336.2\,TNU), assuming an IBD selection with 82.2\% efficiency and an average reactor duty cycle of 11/12 to account for refueling outages. The energy spectrum of reactor antineutrinos extends well beyond the geoneutrino end point, reaching up to about 12\,MeV. For geoneutrino studies, a prompt energy window up to 2.6\,MeV is typically used to fully capture the signal while accounting for detector resolution. Within this window, the corresponding background is 10.7 events per day (332.5\,TNU). The spectral differences between reactor and geoneutrinos, as well as the information in the spectrum above the geoneutrino end point, provide strong leverage to constrain the reactor contribution and isolate the geoneutrino signal with high precision. {In this work, we use the existing constraints on the unoscillated reactor antineutrino spectral shape from the Daya Bay experiment~\cite{DayaBay:2021dqj}, supplemented by an additional uncertainty to account for potential fine-structure effects not resolved by Daya Bay due to its lower energy resolution. This uncertainty, which amounts to about 2\% for most of the energy region, is estimated from a comparison between the Daya Bay measurement and a summation model~\cite{Estienne:2019ujo}. We do not apply the TAO-based constraints used in Refs.~\cite{NMO-paper,JUNO_Precision}, since a geoneutrino measurement can be achieved with significantly lower statistics than required for the neutrino mass ordering measurement, at which point TAO results may not yet be fully available. Moreover, while TAO constraints are important for JUNO’s measurement of the mass ordering and the $\Delta m^2_{31}$ oscillation parameter, we find that geoneutrino sensitivities are only marginally improved by their inclusion, even at higher exposures, owing to the smooth and relatively broad nature of the geoneutrino spectrum.}

\subsubsection{World reactors beyond 300\,km}
\label{subsec:worldrea}

{\it World reactors} refer to reactor antineutrinos originating from commercial reactors located more than 300\,km from the JUNO site. In this analysis, their contribution has been updated using the latest available data from the International Atomic Energy Agency (IAEA). The Power Reactor Information System (PRIS)~\cite{PRIS} provides annually data on reactors worldwide, including their nominal thermal power and monthly load factors.  Using this information, energy spectra per fission of $^{235}$U, $^{239}$Pu, and $^{241}$Pu is obtained, using for each reactor the same flux parametrization and core composition as for the nearby reactors described above. 

We observe a steady increase of the world reactor background. Since 2019 used in~\cite{NMO-paper, JUNO_Precision}, in 2024 the calculated rate is about 30\% higher, mainly due to new reactors in China. Consequently, the world reactor rate has been updated to the latest calculation of 42.2\,TNU, corresponding to 1.4 IBD events per day after IBD selection in JUNO. To account for the unknown yearly increase, the rate uncertainty has been increased to 10\%.
Although the shape of the yearly spectra changes, it was found that updating it has no impact on the geoneutrino measurement, when in the analysis its rate is constrained to the expected value. Therefore, we use the same 5\% shape uncertainty as in~\cite{NMO-paper, JUNO_Precision}.

\subsection{Other Backgrounds}
\label{subsec:OtherBgs}

For this analysis, the non-reactor backgrounds—excluding the updated $^{13}$C($\alpha$,\,n)$^{16}$O background—are the same as those used in the neutrino mass ordering~\cite{NMO-paper} and neutrino oscillation parameters~\cite{JUNO_Precision} sensitivity studies. All backgrounds are summarized in Table~\ref{tab:event_rates}.

After reactor antineutrinos, the most significant background is the accidental background, {formed by the coincidence of two uncorrelated events. Its expected rate of 0.8 events per day is estimated from radioactivity measurements of the various components of the JUNO detector~\cite{JUNO:rad-control}. A 1 \% uncertainty is assigned to the overall rate, while the shape uncertainty is very small due to the precise determination of the energy spectrum using an off-window method, whereby an artificial delayed window, shifted by an arbitrarily long time after the prompt candidate, is used to form uncorrelated pairs.  

An effective muon veto system is expected to suppress the cosmogenic $^9$Li/$^8$He background to the same level as the accidental background, approximately 0.8 events per day. This background can also be reliably estimated using independent data samples in real data. In particular, the identification of IBD-like coincidences that are spatially and temporally correlated with cosmic muons enables the evaluation of untagged $\beta$-neutron decays from $^9$Li/$^8$He. 
}

IBD-like background events from atmospheric neutrinos, fast cosmogenic neutrons, and the $^{13}$C($\alpha$,n)$^{16}$O reaction are expected to occur at significantly lower rates. These contributions carry larger uncertainties, as they typically cannot be fully constrained by data and must instead be estimated through Monte Carlo simulations informed by experimental inputs.

The $^{13}$C($\alpha$,\,n)$^{16}$O background has been recently revised~\cite{JUNOalphan} using state-of-the-art simulations of the interaction. Alpha particles emitted by intrinsic contaminants such as $^{238}$U, $^{232}$Th, and $^{210}$Pb/$^{210}$Po can be captured on $^{13}$C nuclei, resulting in the emission of MeV-scale neutrons. A prompt signal can arise before the delayed neutron capture from neutron elastic scattering on protons, de-excitation $\gamma$ rays emitted by $^{16}$O nuclei produced in excited states, or inelastic interactions of the outgoing neutron with $^{12}$C, yielding a signature that mimics that of an IBD signal.
The open-source SaG4n software~\cite{Men20} was used to calculate the energy depositions from the neutron and any associated $^{16}$O de-excitation products. The detector response was simulated with JUNO's Geant4-based full simulation framework. After application of the IBD selection criteria described in Section~\ref{subsec:IBDselection}, the expected rate is $0.090$ events per day, with 25\% uncertainties in both rate and shape. This update nearly doubles the previous expectation reported in JUNO studies~\cite{NMO-paper, JUNO_Precision}, as it now accounts for unsupported $^{210}$Po, and simultaneously reduces the associated uncertainties by about a factor of two. Given the expected levels of $\alpha$ contamination in the LS and the constraints applied to this background during the fit, this update has a minimal impact on the geoneutrino measurement in the present study.
In the future, the $\alpha$ contamination will be assessed directly from data, providing a means to constrain this background.

\begin{table} [t]
    \centering
    \begin{tabular}{lccccc}
    \hline \hline
    Component & \multicolumn{2}{c}{Rate} & Rate Unc. & Shape Unc. \\
    & $R$ & {$R^{\mathrm geo}$} & $\sigma^{\mathrm{rate}}$ & $\sigma^{\mathrm{shape}}$ \\
              & Total & $E_p$$\,<\,$2.6\,MeV & & \\
              & [day$^{-1}$] & [day$^{-1}$] & [\%] & [\%] \\
    \hline \hline
    Geoneutrinos ($^{238}$U + $^{232}$Th) & 0.92–1.95 &  0.92–1.95 & - & 5 \\
    Geoneutrinos ($^{238}$U)              & 0.71–1.50 & 0.71–1.50 & - & 5 \\
    Geoneutrinos ($^{232}$Th)             & 0.21–0.45 & 0.21–0.45 & - & 5 \\
    \hline
    Reactor antineutrinos                 & 43.2 & 10.7 & -   & Daya Bay~\cite{DayaBay:2021dqj} \\
    World reactors                        & 1.4    & 0.4   & 10  & 5 \\
    Accidentals                           & 0.8    & 0.7   & 1   & 0 \\
    $^9$Li/$^8$He                         & 0.8    & 0.1   & 20  & 10 \\
    Atmospheric neutrinos                 & 0.16   & 0.03  & 50  & 50 \\
    Fast neutrons                         & 0.1    & 0.02  & 100 & 20 \\
    $^{13}$C($\alpha$,\,n)$^{16}$O        & 0.090  & 0.063 & 25  & 25 \\
    \hline \hline
\end{tabular}

    \caption{Expected event rates after IBD selection including muon veto and FV cut (Section~\ref{subsec:IBDselection}) for geoneutrinos and the different backgrounds. For the former, the quoted interval represents the 1$\sigma$ band covering all models from Table~\ref{tab:signals_tot}, e.g. 28.6 to 60.3\,TNU. Individual $^{238}$U and $^{232}$Th contributions are calculated assuming the chondritic mass ratio. For backgrounds, we indicate the total rate as well as the rate within the geoneutrino window with prompt energy $E_p$ below 2.6\,MeV. The last two columns show the rate and shape uncertainties, respectively, applied in this analysis.}
    \label{tab:event_rates}
\end{table}

\section{Analysis Strategy}
\label{sec:analysis}

This section presents the analysis strategy employed to study JUNO's sensitivity to geoneutrinos. Section~\ref{subsec:inputs} outlines the construction of input spectral shapes and the incorporation of the detector response. The fitting procedure, based on a binned $\chi^2$ method utilizing either a covariance matrix or pull terms, is described in Section~\ref{subsec:fitting}, along with the treatment of systematic uncertainties. To assess the sensitivities and validate the robustness of the results, both Asimov and toy pseudo-experiment datasets are used, as detailed in Section~\ref{subsec:StatAnalysis}. The strategy to assess JUNO's discovery potential for the mantle geoneutrino signal is presented in Section~\ref{subsec:mantle_strategy}. 

\subsection{Input Spectral Shapes}
\label{subsec:inputs}

The prediction of the reconstructed prompt energy spectra of the geoneutrino IBD signal and the reactor antineutrino background is carried out using two complementary approaches. The first relies on analytical parameterizations of the IBD reaction kinematics, the liquid scintillator non-linearity, and JUNO’s energy resolution~\cite{JUNOResol}, while the second employs large-scale datasets produced with JUNO’s full simulation framework. The former approach is the same as used in previous sensitivity studies~\cite{NMO-paper,JUNO_Precision}, and it reproduces the same spectra. For geoneutrinos, the contributions from $^{238}$U and $^{232}$Th were simulated separately, and both were found to be in excellent agreement with the analytical spectra. 
{It is worth noting that our analysis accounts for possible spectral distortions in the geoneutrino spectrum arising from neutrino oscillations and the heterogeneous distribution of HPEs by assuming a 5\% shape uncertainty, which encompasses these effects.} For the extraction of the TNU geoneutrino signal in real data analysis, only the spectral deformation is relevant. The effect of oscillation on the expected event rate is, however, included in the TNU signal prediction of different models, as described in more detail in Appendix. An average constant survival probability of 55\% is assumed for the mantle, while for the lithosphere we use directly the TNU values predicted by the authors of each model. Reactor neutrinos were simulated unoscillated, as the oscillated spectra can be obtained by applying the survival probability. Small differences with respect to the analytical spectra were observed but found to have a negligible impact on the resulting sensitivities. The remaining background spectra are modeled with the analytical approach. The accidental spectrum is derived from the latest characterization of detector materials, incorporating the energy, position, and rate of uncorrelated radioactive events~\cite{JUNO:rad-control}. The spectrum of cosmogenic $^{9}$Li–$^{8}$He is adopted from the calculation benchmarked at Daya Bay~\cite{DYB2016}, and the fast neutron contribution is taken to be flat. The ``world reactors" spectrum, arising from cores located more than 300\,km from JUNO, has been discussed in Section~\ref{sec:SignalBackground}. Except for the $(\alpha,n)$ background, whose rate and spectral shape have been updated with a dedicated Monte Carlo simulation (see Section~\ref{subsec:OtherBgs}), all of these spectra are unchanged from previous analyses.

\begin{figure}[t]
    \centering
    \subfloat[]{\includegraphics[width=0.45\linewidth]{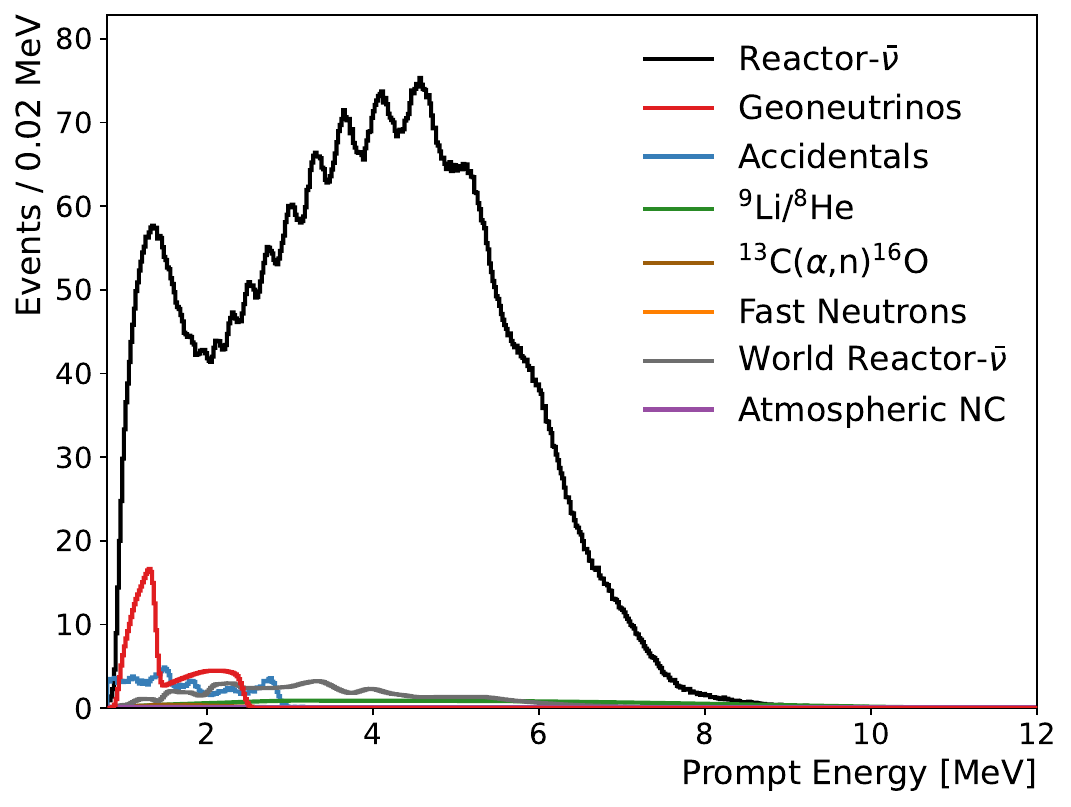}}
    \subfloat[]{\includegraphics[width=0.45\linewidth]{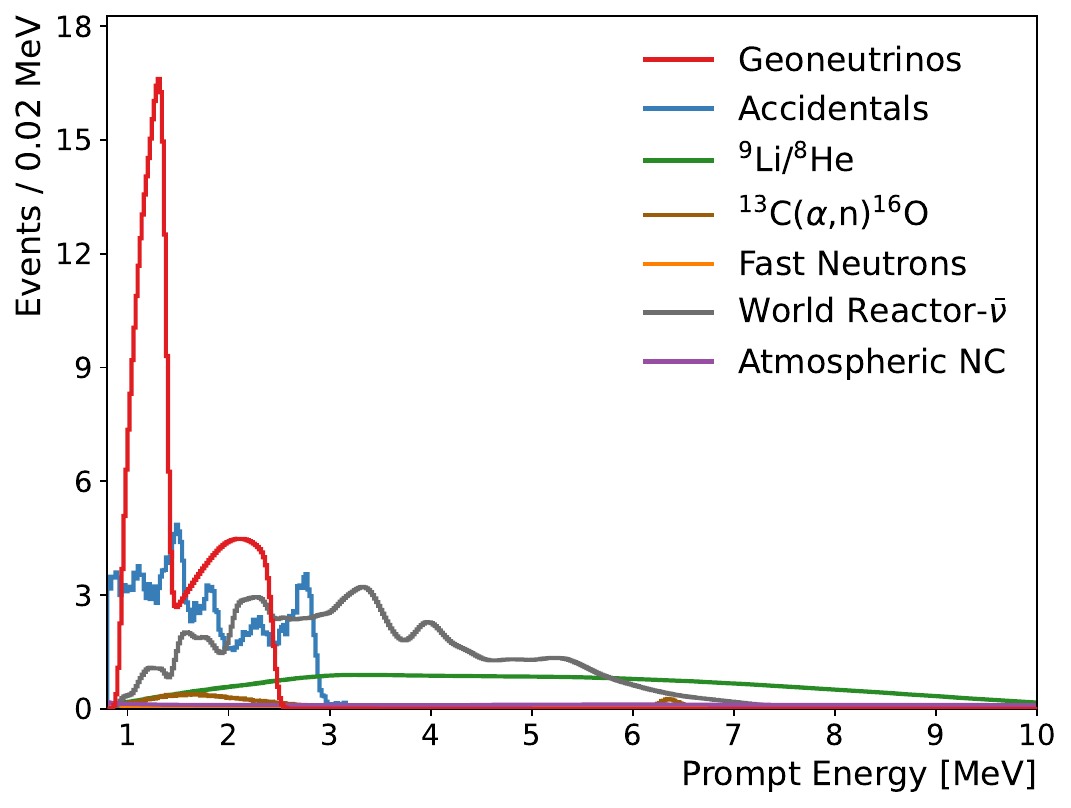}}
    \caption{ (a) Spectral shapes of geoneutrinos, reactor neutrinos, and other backgrounds normalized to 1\,year of exposure using the expected rates from Table~\ref{tab:event_rates}, with 40\,TNU for geoneutrino signal. Total geoneutrino spectrum shown in red corresponds to  $^{238}$U and $^{232}$Th components fixed according to chondritic ratio. (b) A zoomed-in version of (a) with reactor spectrum removed in order to better appreciate comparison of geoneutrino spectra with different backgrounds.}
    \label{fig:sig_bkgs}
\end{figure}

The resulting spectral shapes of geoneutrinos, reactor neutrinos, and other backgrounds, normalized to their expected rates under 1\,year of exposure as given in Table~\ref{tab:event_rates} and for a 40\,TNU geoneutrino signal, are shown in Fig.~\ref{fig:sig_bkgs}. We adopt the values of the oscillation parameters from PDG~2024~\cite{PDG2024} as nominal.

\subsection{Fitting Procedure}
\label{subsec:fitting}

We perform our sensitivity studies using the binned $\chi^2$ method. We fit the energy spectrum of the prompt IBD candidates from 0.8\,MeV to 12.0\,MeV, and adopt a differential binning strategy to avoid bins with zero or very low bin content in realistic datasets. Specifically, the region between 1 and 7 MeV is divided into 20\,keV bins, while the lower and higher energy regions have bin widths that range between 0.2 and 2\,MeV, such that the probability of having a fluctuation to zero with one year of exposure is less than 0.02\%.  The fits are performed using a total of 314 energy bins and with the iMINUIT package in Python~\cite{dembinski_2025_15157028, JAMES1975343}.

The combined Neyman-Pearson (CNP) $\chi^2$ is used for minimization as it is found to have the least bias with low statistics compared to other test statistics~\cite{CNP}. The basic CNP $\chi^2$ that only includes statistical uncertainty can be written as:
\begin{equation}\label{eq:CNP}
    \chi^2_{\textrm{CNP}} = \sum_{i=1}^\textrm{Nbins}\frac{(T_i(\bm{\theta}) - M_i)^2}{3\big/\left(\frac{1}{M_i} + \frac{2}{T_i(\bm{\theta})}\right)}.
\end{equation}
Here, $M_i$ represents the measured number of events. The template model number of events, $T_i(\bm{\theta})$, depends on the 
vector of fit parameters $\bm{\theta}$ and is the sum of geoneutrinos, reactor antineutrinos, and other background events, in the $i$-th energy bin. Fit parameters include the oscillation parameters $\sin^2\theta_{12}$, $\Delta m^2_{21}$, $\Delta m^2_{31}$, and $\sin^2\theta_{13}$, the number of reactor antineutrinos $N_{\text{rea}}$, and the number of geoneutrinos $N_{\text{geo}}$. The latter parameter is used when Th/U ratio is fixed in the fit, while parameters $N_{\text{U}}$ and $N_{\text{Th}}$ are used instead of $N_{\text{geo}}$ when the $^{238}$U and $^{232}$Th components are fit independently. {All the oscillation parameters are left free in the fit without constraints. Applying the Daya Bay constraint on $\sin^2 \theta_{13}$~\cite{DayaBay:2022orm}, which is more precise than any JUNO measurement, has very little impact on the sensitivities.
}

The systematic uncertainties considered in the fit include the rate ($\sigma^{\mathrm{rate}}$) and spectral shape ($\sigma^{\mathrm{shape}}$) of the different components, as listed in Table~\ref{tab:event_rates}. The non-linearity of the liquid scintillator energy scale for positrons is also considered following the same approach described in Refs.~\cite{NMO-paper, JUNO_Precision}, under the assumption that a calibration precision comparable to that achieved at Daya Bay can be obtained. In this approach, four representative curves corresponding to typical one-sigma variations are randomly weighted to span the full uncertainty range. Other effects such as the uncertainties on the non-equilibrium and spent nuclear fuel corrections applied on the reactor spectrum, the error on JUNO's energy resolution, and the uncertainty on the Earth's matter density used in the neutrino oscillation formula, were found to have a negligible impact and are thus not considered.

The uncertainties are treated through both covariance matrices and pull terms. In general, the two approaches yield consistent results, and any small residual differences are included in the systematic uncertainty of the final result. In the covariance matrix approach, the CNP $\chi^2$ with all sources of uncertainty becomes:
\begin{equation}\label{eq:CNP-CM}
    \chi^2_{\textrm{CNP-CM}} = \sum_{i,j=1}^{N_\textrm{bins}} (M_{i}- 
    T_{i}(\bm{\theta}))
    \cdot V^{-1}_{ij} \cdot (M_{j}- T_{j}(\bm{\theta})),
\end{equation}
where $i$ and $j$ are represent the energy bins in a two-dimensional covariance matrix  $V_{ij}$. We note that, in the statistics-only case, $V_{ij} = V^{\text{stat}}_{ij}$ is diagonal and becomes identical to the denominator in Eq.~\eqref{eq:CNP}, 
 \begin{equation}
     V^{\text{stat}}_{ij} = 3\big/\left(\frac{1}{M_i} + \frac{2}{T_i(\bm{\theta})}\right), 
 \end{equation}
thus showing the equivalence between the two approaches. When systematic uncertainties are included, the systematic covariance matrix 
$V^{\text{syst}}_{ij}$ is constructed by summing the individual systematic uncertainty matrices that include the rate and shape uncertainties of the spectral components and the non-linearity effect, and the total covariance is given by $V_{ij} = V^{\text{stat}}_{ij} +  V^{\text{syst}}_{ij}$. The shape uFncertainties of all the spectral components  $\sigma_{\mathrm{sp,i}}^{\mathrm{shape}}$
are modeled as bin-to-bin uncorrelated and the rate uncertainties of the backgrounds $\sigma^{\mathrm{rate}}_{\mathrm{b}}$ as bin-to-bin correlated. {The uncertainties for each spectral component were estimated individually using the methods explained in Section~\ref{sec:SignalBackground}.} The non-linearity uncertainty is propagated through the detector response model parameters, and the corresponding covariance matrix  is obtained by generating 10,000 ($N_{\textrm{exps}}$) fluctuated spectra with the fluctuated non-linearity curves and computing the difference between the nominal spectrum and the fluctuated spectra. The complete systematic covariance matrix can thus be written as
\begin{equation}
    V^{\text{syst}}_{ij} = \sum_{sp=1}^{N_\textrm{spectra}} (\sigma^{\mathrm{shape}}_{\textrm{sp}, i})^2 M_i \cdot M_i + \sum_{b=1}^{N_\textrm{bckgs}} (\sigma^\textrm{rate}_{\textrm{b},i})^2 M_i \cdot M_j +  \frac{1}{N_\textrm{exps}} \sum_{k=1}^{N_\textrm{exps}}(M_i - x_i^k) (M_j - x_j^k),
\end{equation}
where $x^k$ represents the bin content of the $k$-th fluctuated spectrum.

In the pull-term approach, the $\chi^2$ is written as:

\begin{equation}
\chi^2_{\textrm{CNP-pull}} = 
\sum_{i=1}^{N_\textrm{bins}}
\frac{(T_i(\bm{\theta}) - M_i)^2}
{3\Big/\left(\displaystyle\frac{1}{M_i} + \frac{2}{T_i(\bm{\theta}) + \left(\sum_{sp=1}^{N_\textrm{spectra}} T_i(\bm{\theta}) \cdot \sigma^{\mathrm{shape}}_{\mathrm{sp},i}\right)^2}\right)} +
\sum_{b=1}^{N_\textrm{bckgs}} \left(\frac{T_{\mathrm {b}}(\bm{\theta}) - \bar{\eta}_{\mathrm{b}}}{\sigma_{\mathrm {b}}^{\mathrm{rate}}}\right)^2 +
\sum_{l=1}^{4} \alpha_l^2.
\end{equation}
Here, the shape uncertainties $\sigma^{\mathrm{shape}}$ (Table~\ref{tab:event_rates}) of all components are added in quadrature with the statistical uncertainties in each bin. The fitted number of events for each background $b$, $T^b(\bm{\theta})$, is constrained with a penalty term, where $\bar{\eta}_b$ denotes the expected number of events and $\sigma_{b}^{\mathrm{rate}}$ the associated rate uncertainty. In addition, four nuisance parameters $\alpha_{l}$ centered at 0 with unit standard deviation are introduced to scale the four pull curves used to model the non-linearity uncertainty~\cite{NMO-paper}, and their contributions are added in quadrature to the $\chi^2$ as penalty terms.

\begin{figure}[t]
    \centering
    {Geoneutrino Signal: 20 TNU, Exposure: 1 year}\\[-0.5em]
    \subfloat[]{\includegraphics[width=0.45\linewidth]{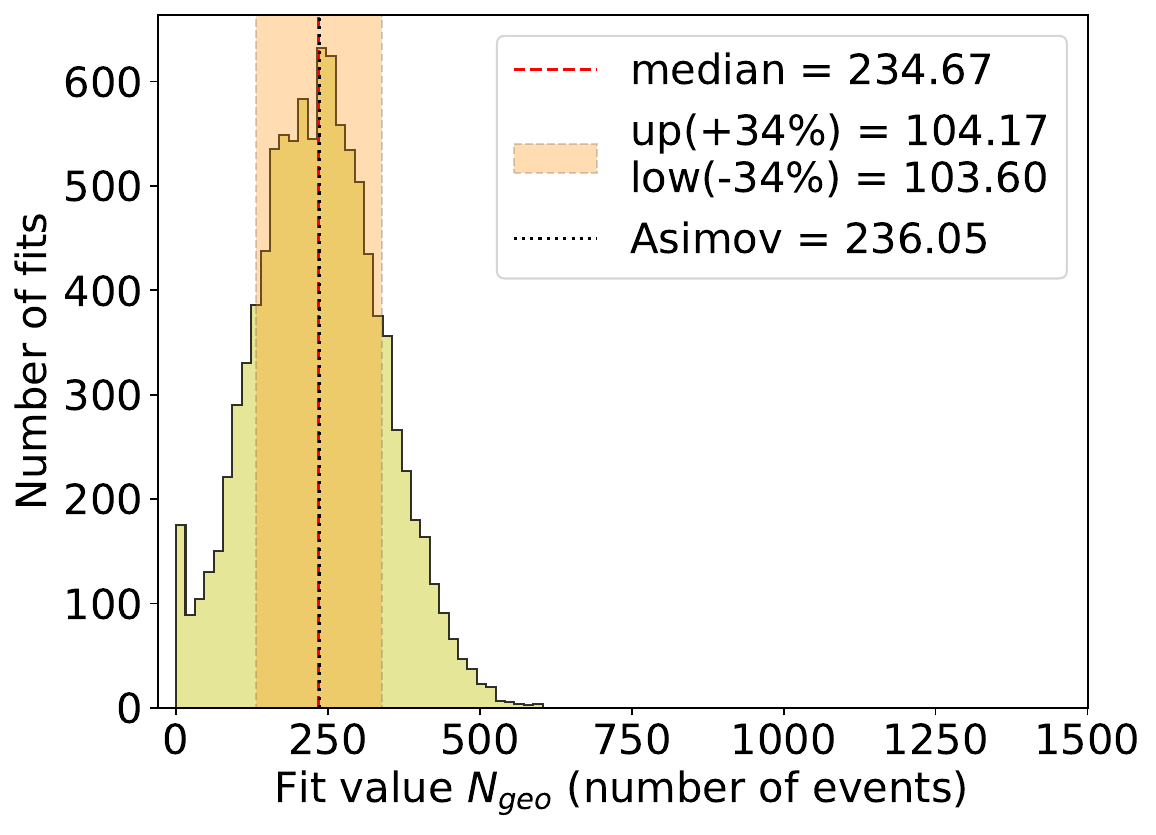}}%
    \subfloat[]{\includegraphics[width=0.45\linewidth]{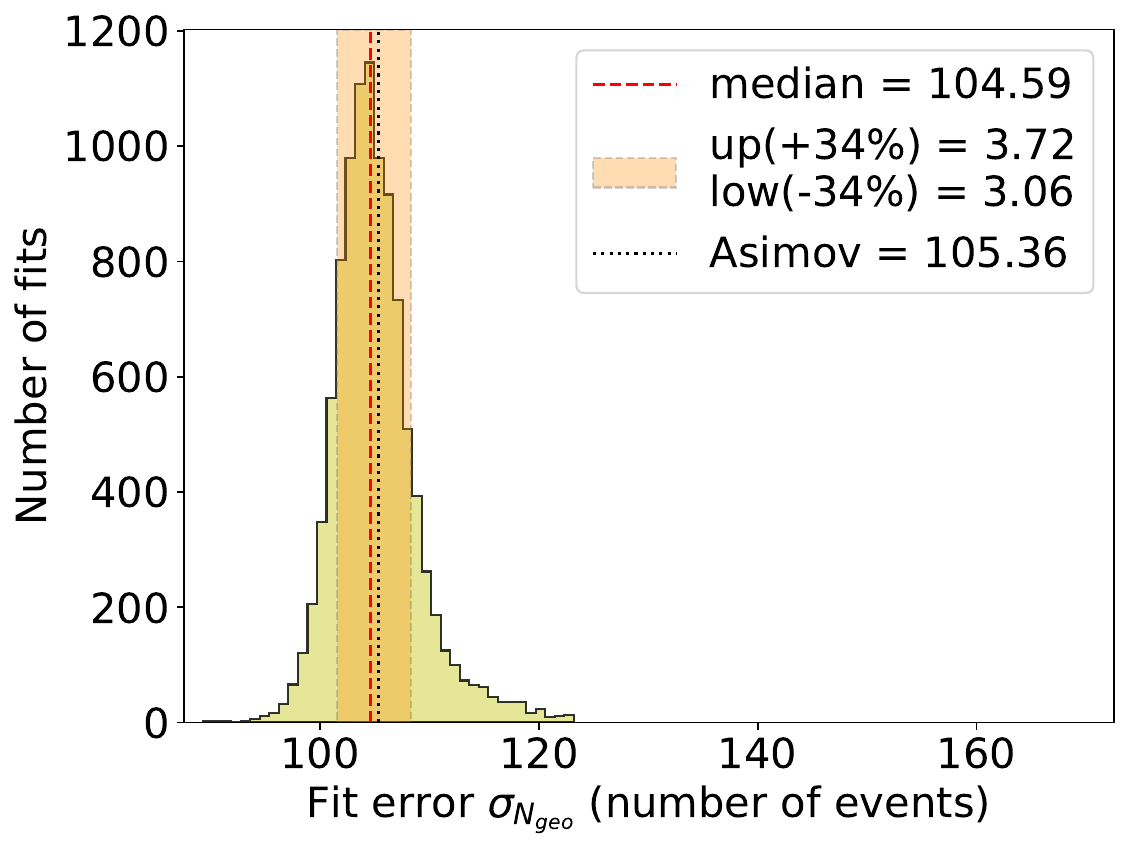}}\\
    \subfloat[]{\includegraphics[width=0.45\linewidth]{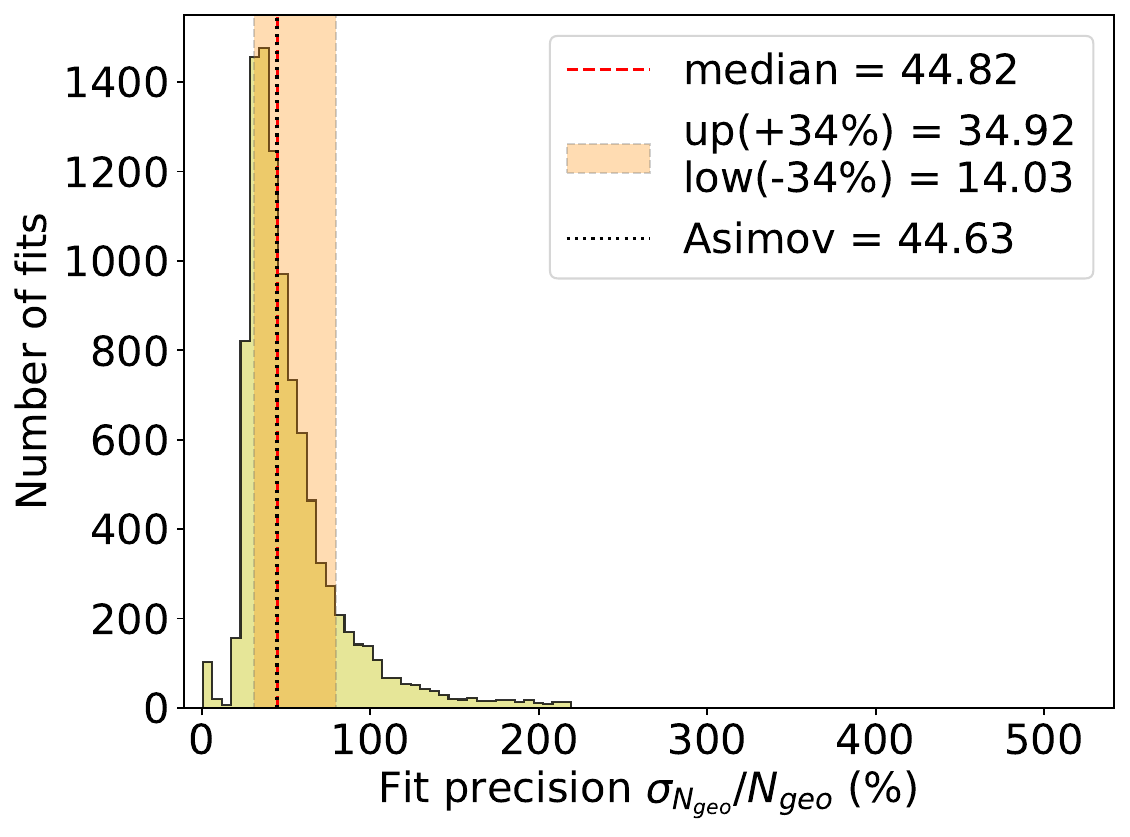}}
    \caption{Distributions of fit results from 10,000 toy MC pseudo-datasets are shown and compared to the Asimov approach results (dotted vertical lines) for 20\,TNU geoneutrino signal and 1 year exposure. The dashed lines indicate the medians of the distributions, while the shaded vertical bands correspond to the $\pm34\%$ confidence intervals. (a) Fitted number of geoneutrino events, $N_{\mathrm{geo}}$. A clear peak in $N_{\mathrm{geo}}$ converging to the physical boundary at zero is observed. (b) Corresponding fit uncertainty $\sigma_{N_{\mathrm{geo}}}$. (c) Relative precision, defined as $\sigma_{N_{\mathrm{geo}}} / N_{\mathrm{geo}}$, represents the JUNO sensitivity to measure geoneutrinos.}
\label{fig:ToyMC_demonstration}    
\end{figure}

\subsection{Statistical Approaches}
\label{subsec:StatAnalysis}

The framework described above supports fitting both Asimov datasets, which represent the nominal expectation without fluctuations, and pseudo-experiment (toy MC) datasets, which include statistical and systematic variations. The Asimov approach provides an estimate of the expected sensitivity from a single dataset, whereas the toy MC method employs 10,000 pseudo-datasets to extract distributions of best-fit values and uncertainties, thereby determining the achievable precision from the median. Both methods, using either a covariance matrix or pull terms as described in the previous subsection, are applied to the Asimov and toy MC datasets. The results are in excellent agreement, with deviations between the Asimov and toy MC approaches observed at lower exposures or reduced signal rates, as expected. The minor differences between the covariance and pull-term approaches are treated as a source of systematic uncertainty.

For the Asimov case, each input spectrum is scaled according to the expected number of events, based on the assumed experimental exposure and the rates provided in Table~\ref{tab:event_rates}. These individual spectra are then summed to produce a single, non-fluctuated spectrum, with error bars reflecting realistic statistical uncertainties. It is worth noting that when using shapes generated from Monte Carlo simulations, the simulated statistics are significantly higher than the expected number of events. 

\begin{figure}[t]
    \centering
    \centering
    \subfloat[]{\includegraphics[width=0.5\linewidth]{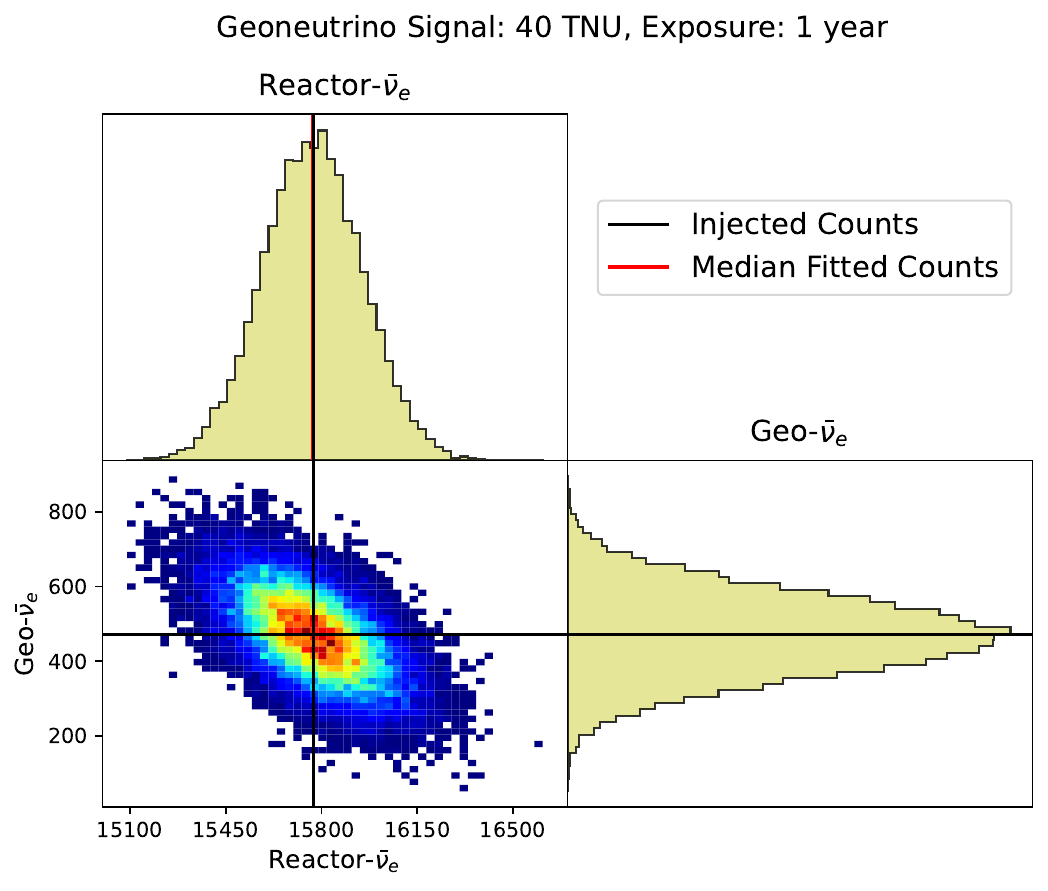}}
    \subfloat[]{\includegraphics[width=0.5\linewidth]{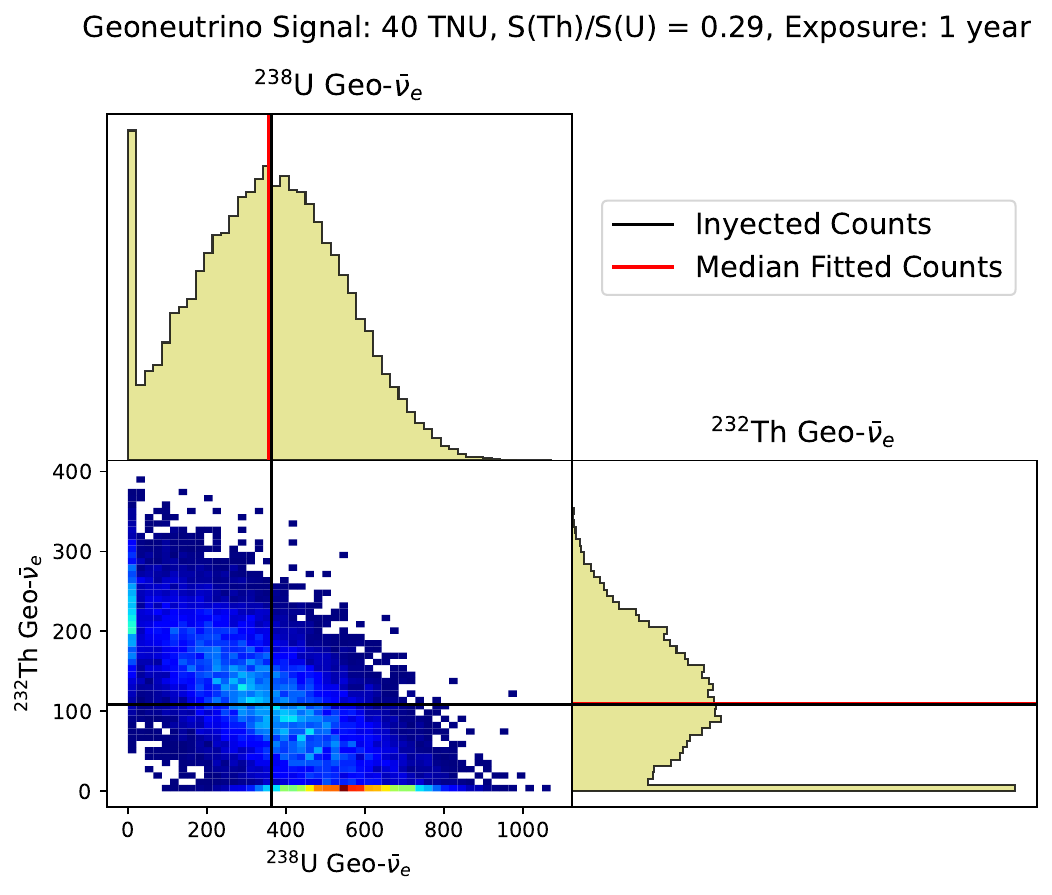}}
    \caption{Fit result distributions from 10,000 toy MC pseudo-datasets, showing spectral component correlations for one year of exposure and a 40\,TNU geoneutrino signal. Left: anti-correlation between the geoneutrino component $N_{\mathrm{geo}}$ and the reactor antineutrino component $N_{\mathrm{rea}}$, assuming a fixed Th/U ratio for geoneutrinos at the chondritic value. Right: anti-correlation between the individual $N_{\mathrm{U}}$ and $N_{\mathrm{Th}}$ components.}
    \label{fig:correlations}
\end{figure}

For each toy MC dataset, the systematic uncertainties are varied by independently sampling each source from a Gaussian distribution with mean equal to the central expected value and standard deviation given by the corresponding uncertainty, avoiding over-constraints from highly correlated parameters. 
The toy MC approach enables estimating the probability that the geoneutrino signal cannot be identified and converges to the physical boundary at zero, which is particularly relevant at low exposures and small signal rates. This is illustrated for a 20\,TNU geoneutrino signal and a 1 year exposure in Fig.~\ref{fig:ToyMC_demonstration} (a). We exclude from further analysis any such configurations where the probability of failing to identify the geoneutrino signal exceeds 0.27\%, which corresponds to a 3$\sigma$ confidence level. 
The remaining panels of Fig.~\ref{fig:ToyMC_demonstration} illustrate the procedure used to evaluate JUNO’s precision in measuring geoneutrinos. This precision is quantified as the median of the distribution of individual fit uncertainties, defined as $\sigma_{N_{\mathrm{geo}}} / N_{\mathrm{geo}}$, with the corresponding uncertainty given by the $\pm 34\%$ confidence interval around the median. Note that these intervals are asymmetric due to the underlying asymmetry in the distribution of fit errors $\sigma_{N_{\mathrm{geo}}}$. The toy MC approach allows also to study correlations among spectral components, as illustrated in Fig.~\ref{fig:correlations}. The adopted method automatically takes such correlations into consideration.
$$H_{ij} = \frac{\partial^2 \chi^2}{\partial \theta_i \partial \theta_j}\Big|_{\text{min}}$$ 

 $$\sigma_{N_{\text{geo}}} = \sqrt{\text{Cov}(N_{\text{geo}}, N_{\text{geo}})} =                           
  \sqrt{(H^{-1}){N{\text{geo}}, N_{\text{geo}}}}$$
$$\lambda = -2\ln\frac{L_{\max}(\text{null})}{L_{\max}(\text{alt})} \xrightarrow{d} \chi^2_{n}$$
$$p\text{-value} = P(\Delta\chi^2_0 \geq \text{median}(\Delta\chi^2_1) | H_0 \text{ true})$$

  $$1 - p\text{-value} = \text{discovery power}$$

\subsection{Mantle Discovery Potential Analysis Strategy}
\label{subsec:mantle_strategy}

This work also includes an assessment of JUNO's ability to distinguish the mantle signal from the lithospheric signal. We evaluated all three Bulk Silicate Earth (BSE) model types—Poor-H, Medium-H, and Rich-H—by injecting their respective signal strengths and thorium-to-uranium ratios (see Table~\ref{tab:mantle_signal_text}). To comprehensively represent current lithospheric models, discussed in Section~\ref{subsec:litho}, we varied the lithospheric signal strength from 25 to 50\,TNU and its associated uncertainty from 5\% to 25\%. 

For each set of mantle and lithospheric signal strengths, we generated two types of datasets, both including the lithospheric signal, but only one also incorporating the mantle contribution. These datasets were used to evaluate our ability to distinguish between the null hypothesis (no mantle signal) and the alternative hypothesis, in which the mantle signal is present. We tested the procedure using both Asimov and large-statistics toy MC datasets and found the resulting discovery potentials to be in good agreement. For computational reasons, the full scan was therefore performed using only the Asimov datasets. 

Each dataset was fitted twice using the procedure described in Section~\ref{subsec:fitting}. The only modification was that, instead of fitting the total geoneutrino signal $N_{\text{geo}}$, we fitted the lithospheric signal $N_{\text{litho}}$ and the mantle signal $N_{\text{mantle}}$ separately. The lithospheric signal was constrained according to the assumed level of prior knowledge, while the mantle signal was either fixed to zero, yielding $\chi^2_0$, or treated as a free parameter, yielding $\chi^2_1$. 
We then define the test statistic $\Delta \chi^2 = \chi^2_0 - \chi^2_1$, with $\Delta \chi^2_0$ for the dataset without an injected mantle signal and $\Delta \chi^2_1$ for the dataset with the mantle signal. In the Asimov approach, $\Delta \chi^2_0 = 0$, since in the absence of statistical fluctuations, the data set without the mantle can be perfectly fit under both hypotheses, yielding $\chi^2_0 = \chi^2_1 = 0$. In the toy MC approach, $\Delta \chi^2_0$ was typically also close to zero, though always positive, as statistical fluctuations can mimic a mantle signal and 
reduce $\chi^2_1$ relative to $\chi^2_0$. In contrast, for datasets including a mantle contribution, 
the test statistic $\Delta \chi^2_1$ can attain large values, 
since the spectral fit is significantly worse when the mantle signal is omitted. In the Asimov approach, $\Delta \chi^2_1 = \chi^2_0$, since $\chi^2_1 = 0$. The discovery potential, expressed in units of $\sigma$, is thus defined as $\sqrt{\chi^2_0}$, obtained from fitting the dataset with an injected mantle signal under the null hypothesis, where the mantle contribution is set to zero.

\begin{figure}[t]
    \centering
    \includegraphics[width=0.55\linewidth]{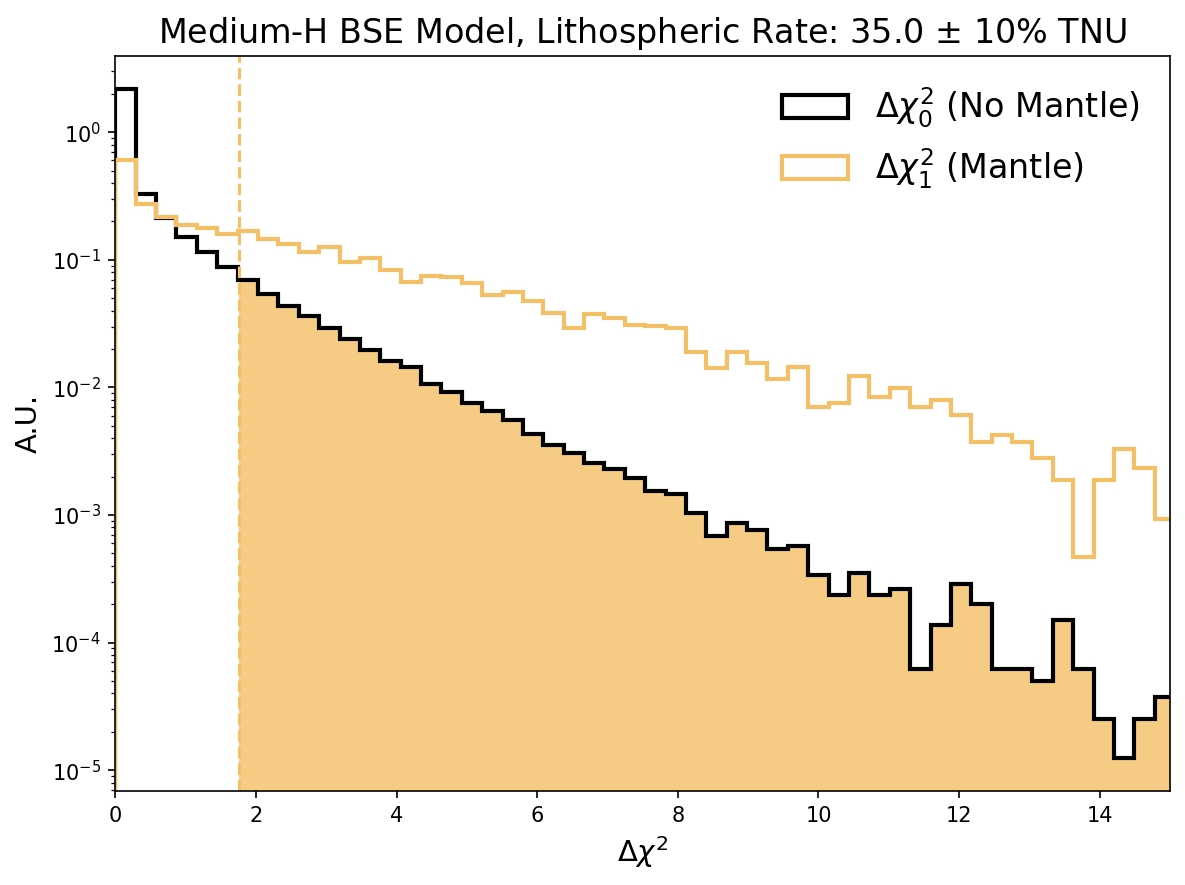}
    \caption{Demonstration of JUNO’s discovery potential for the mantle geoneutrino signal using six years of data with the toy MC approach. We present $\Delta \chi^2_0$ for datasets without an injected mantle signal (black solid line), alongside $\Delta \chi^2_1$ for datasets including mantle signals corresponding to the Medium-H BSE model (solid orange line). The vertical orange dashed line represents median of $\Delta \chi^2_1$ distribution. The discovery potential is defined as the $p$-value, e.g. the probability that $\Delta \chi^2_0$ exceeds the median of the $\Delta \chi^2_1$ distribution, represented by the orange filled area. The injected lithospheric signal is set to 35\,TNU on which a constraint with 10\% uncertainty is applied during the fit. 
}
    \label{fig:mantle_discovery_dchi2}
\end{figure}

Figure~\ref{fig:mantle_discovery_dchi2} illustrates the distributions of $\Delta \chi^2_0$ and $\Delta \chi^2_1$ for the Medium-H BSE model, based on simulated datasets corresponding to six years of exposure using the toy MC approach. The lithospheric signal is set to 35\,TNU, with a 10\% constraint applied during the fit. We define the discovery potential $p$-value as the probability that $\Delta \chi^2_0$ exceeds the median of the $\Delta \chi^2_1$ distribution, expressed by the filled orange area. To express the discovery potential in terms of significance, we convert the $p$-value to an equivalent number of standard deviations ($n\sigma$) assuming a normal distribution.

\section{Results on JUNO Expected Sensitivity to Geoneutrinos}
\label{sec:sensitivity}

The expected precision on the geoneutrino signal at JUNO depends strongly on both the integrated exposure and the true geoneutrino signal strength. In the following sections, we present sensitivity projections for exposures of up to 10 years, covering the range of geoneutrino signals predicted by different geological models as discussed earlier. Section~\ref{sec:ResultsUThFixed} focuses on the measurement of the total geoneutrino signal under the condition of a fixed thorium-to-uranium (Th/U) ratio in the fit, corresponding to the injected chondritic value. Section~\ref{sec:ResultsUThFree} explores the case where the U and Th contributions are fitted independently. Finally, Section~\ref{sec:ResultsMantle} is dedicated to assessing JUNO’s potential to detect the mantle geoneutrino signal as an excess above the constrained lithospheric contribution. All results were independently cross checked by different analyses groups and the eventual sub-dominant differences were treated as an additional systematic error.

\subsection{Total geoneutrino signal with U and Th fixed ratio}
\label{sec:ResultsUThFixed}

\begin{figure}[t]
    \centering
    \includegraphics[width=0.7\linewidth]{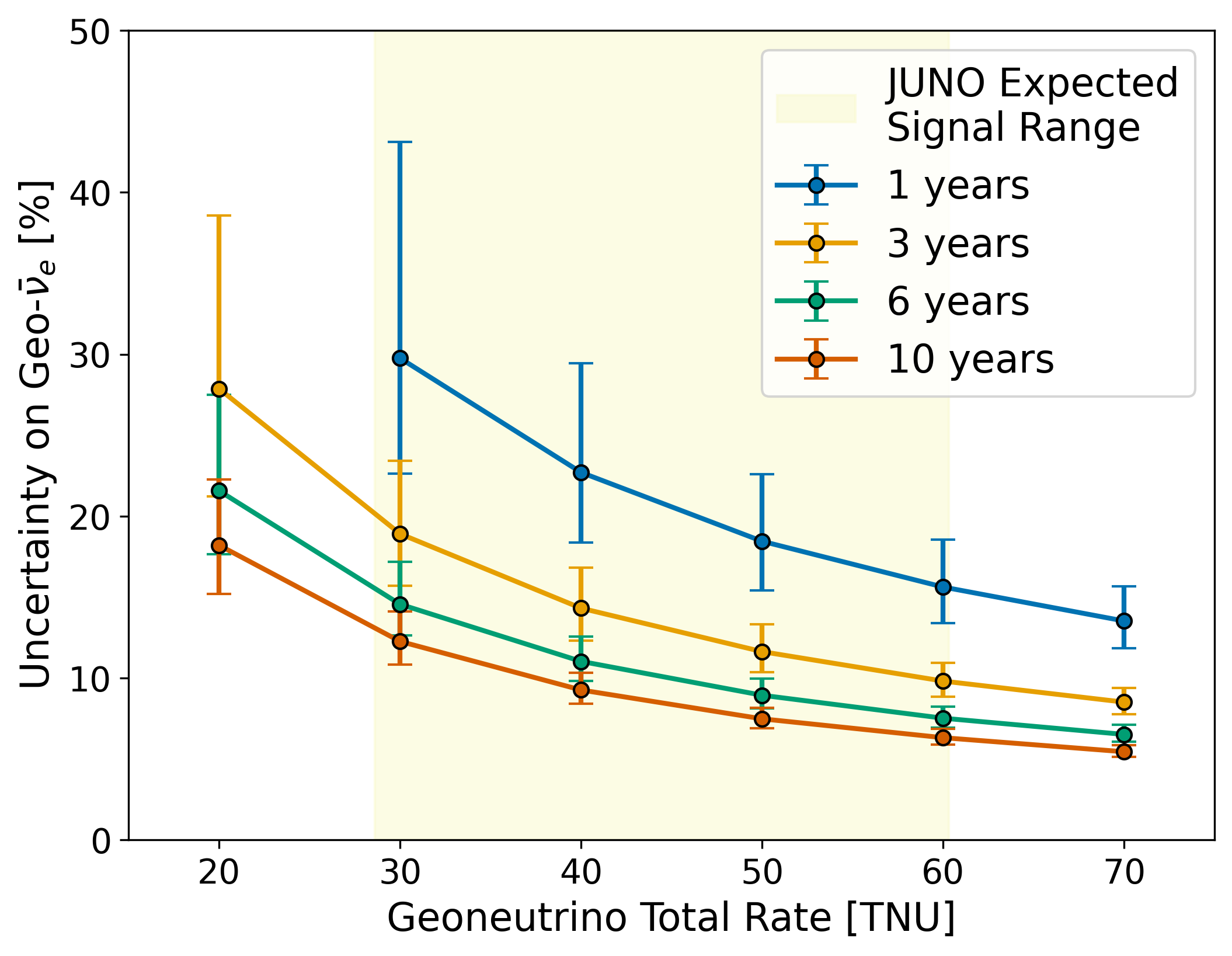}
    \caption{Expected JUNO sensitivity to the total geoneutrino signal, assuming a fixed chondritic Th/U ratio, as a function of the total geoneutrino signal. The solid blue, orange, green, and red curves represent exposures of 1, 3, 6, and 10 years, respectively. The vertical yellow shaded area indicates the 1$\sigma$ range from Table~\ref{tab:signals_tot}, encompassing the spread of different BSE models as well as various predictions for the lithospheric contribution.}
    \label{fig:totageo_scan}
\end{figure}

The JUNO projected precision for measuring the total geoneutrino signal, assuming a fixed Th/U ratio in the fit, is shown in Fig.~\ref{fig:totageo_scan}. Results are presented for exposures of 1, 3, 6, and 10 years, and for geoneutrino signal values ranging from 20 to 70~TNU, covering the full range of predictions from various geological models (indicated by the yellow area representing 1$\sigma$ interval from Table~\ref{tab:signals_tot}). For each exposure and signal value, the central value is taken as the average between the two statistical approaches—Asimov and toy MC. The associated uncertainty corresponds to the $\pm34\%$ confidence interval of the distribution of relative precisions obtained from 10{,}000 toy MC fits, as demonstrated in Fig.~\ref{fig:ToyMC_demonstration}(c). These intervals are asymmetric, reflecting the inherently asymmetric nature of the fit uncertainties in individual toy MC realizations, as illustrated in Fig.~\ref{fig:ToyMC_demonstration}(b). We present the estimated JUNO precision only for cases where the probability that the fitted signal differs from zero is sufficiently high, approximately corresponding to a discovery potential greater than 3$\sigma$. Therefore, the case shown in Fig.~\ref{fig:ToyMC_demonstration}, where the peak at 0 fitted geoneutrino signal is clearly visible, is not represented in the results shown in Fig.~\ref{fig:totageo_scan}.

For a low-signal scenario of approximately 30~TNU, corresponding to the lower edge of the 1$\sigma$ band across various geological models, JUNO would achieve a precision of about 18\%, comparable to existing geoneutrino measurements by Borexino and KamLAND, after roughly 3 to 6 years. For an intermediate signal of 40~TNU, falling well in the yellow band of expected geological predictions, the uncertainty improves more significantly: from $22.7^{+6.7}_{-4.3}\%$ after 1 year to $14.3^{+2.5}_{-2.0}\%$ at 3 years, $11.0^{+1.5}_{-1.2}\%$ at 6 years, and $9.3^{+1.1}_{-0.9}\%$ after 10 years. At the high end, assuming a total signal of 60--70~TNU, the uncertainty can be reduced further, reaching values as low as $5.4\%$ in 10 years.

This level of precision will enable JUNO to distinguish between competing models of Earth's radiogenic heat production and to contribute substantially to our understanding of the Earth's internal composition.

\newpage
\subsection{Uranium and Thorium Independent Contributions}
\label{sec:ResultsUThFree}

\begin{figure}[h!]%
    \centering
    \subfloat[Uranium]{\includegraphics[width=0.45\textwidth]{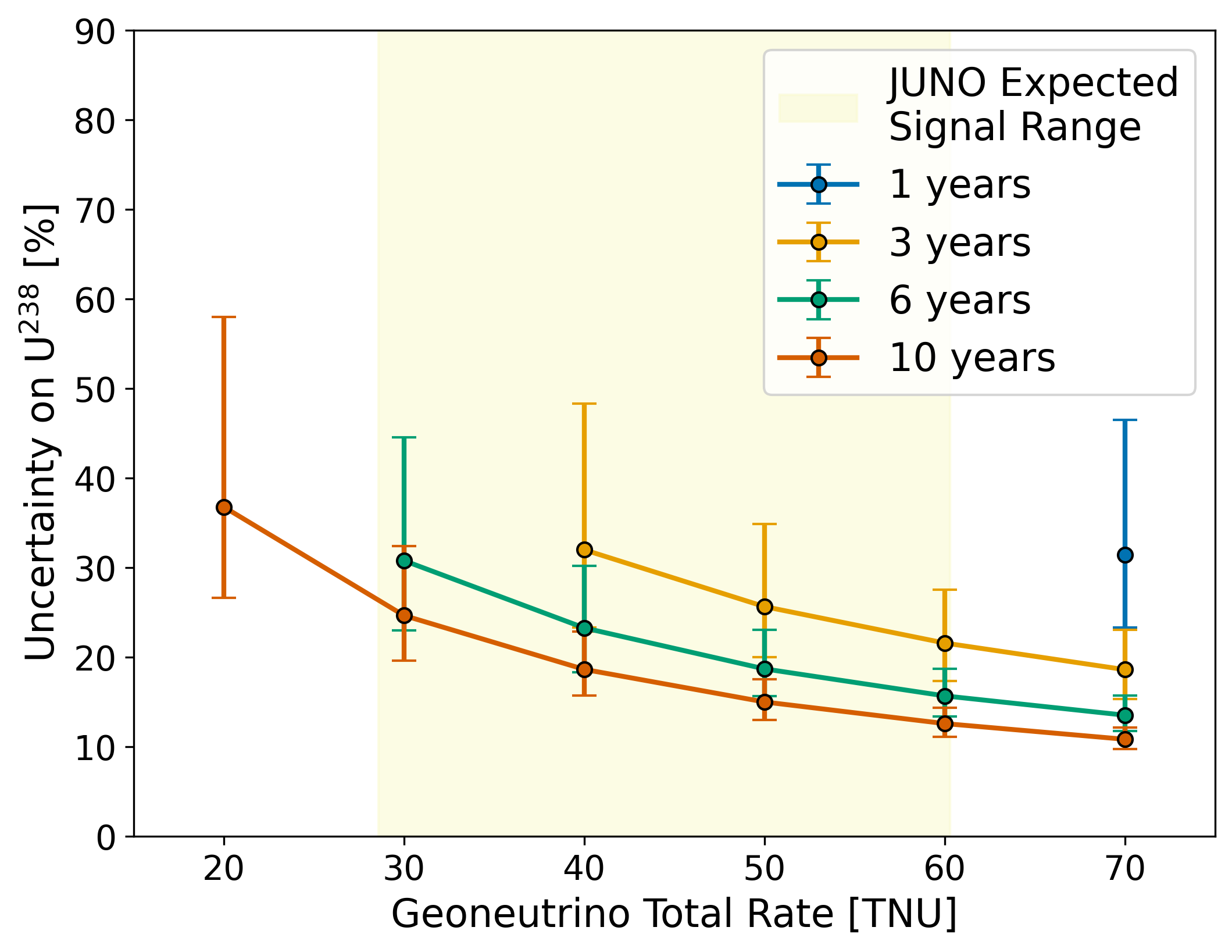}\label{fig:ugeo_scan}}%
    \subfloat[Thorium]{\includegraphics[width=0.45\textwidth]{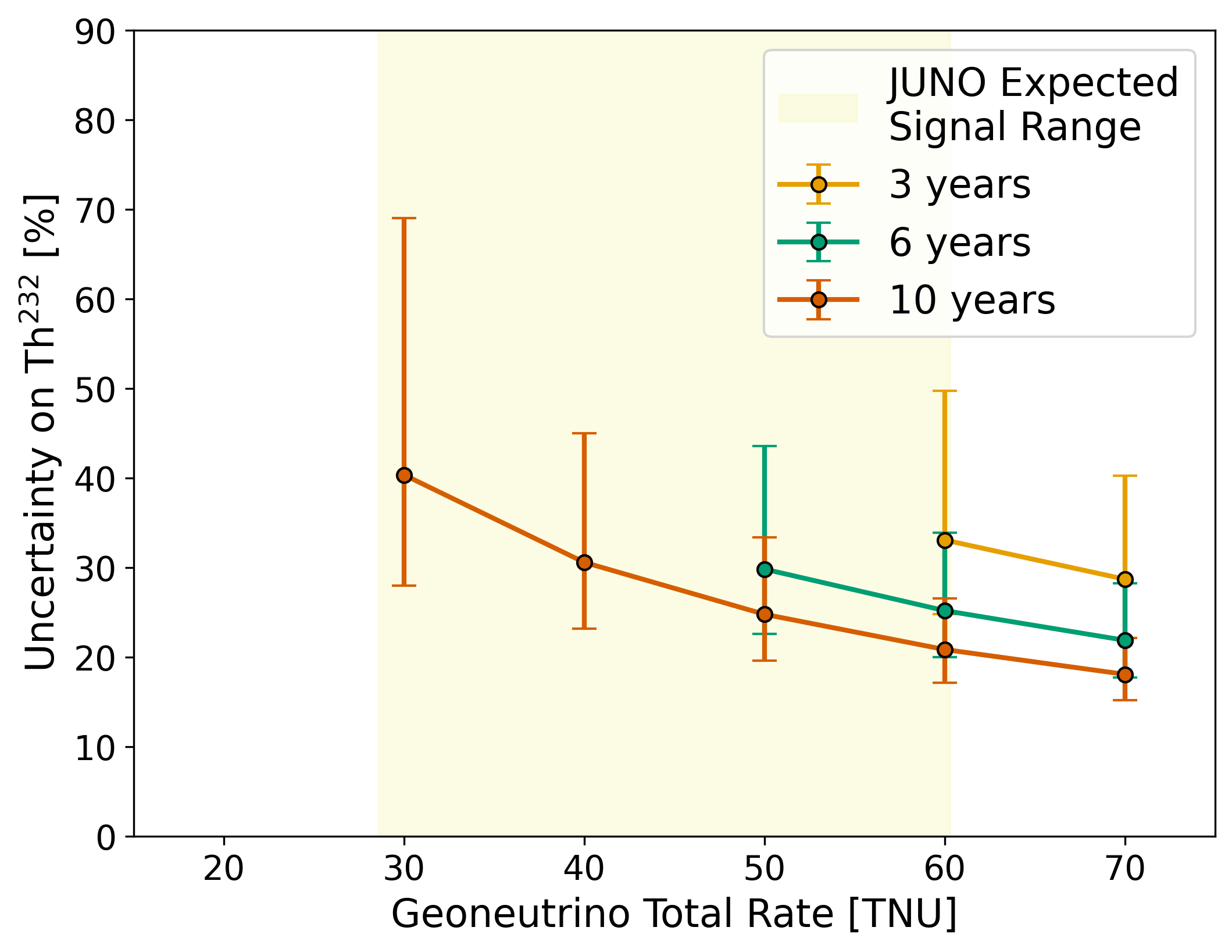}\label{fig:thgeo_scan}}\\
    \subfloat[Uranium + Thorium]{\includegraphics[width=0.45\textwidth]{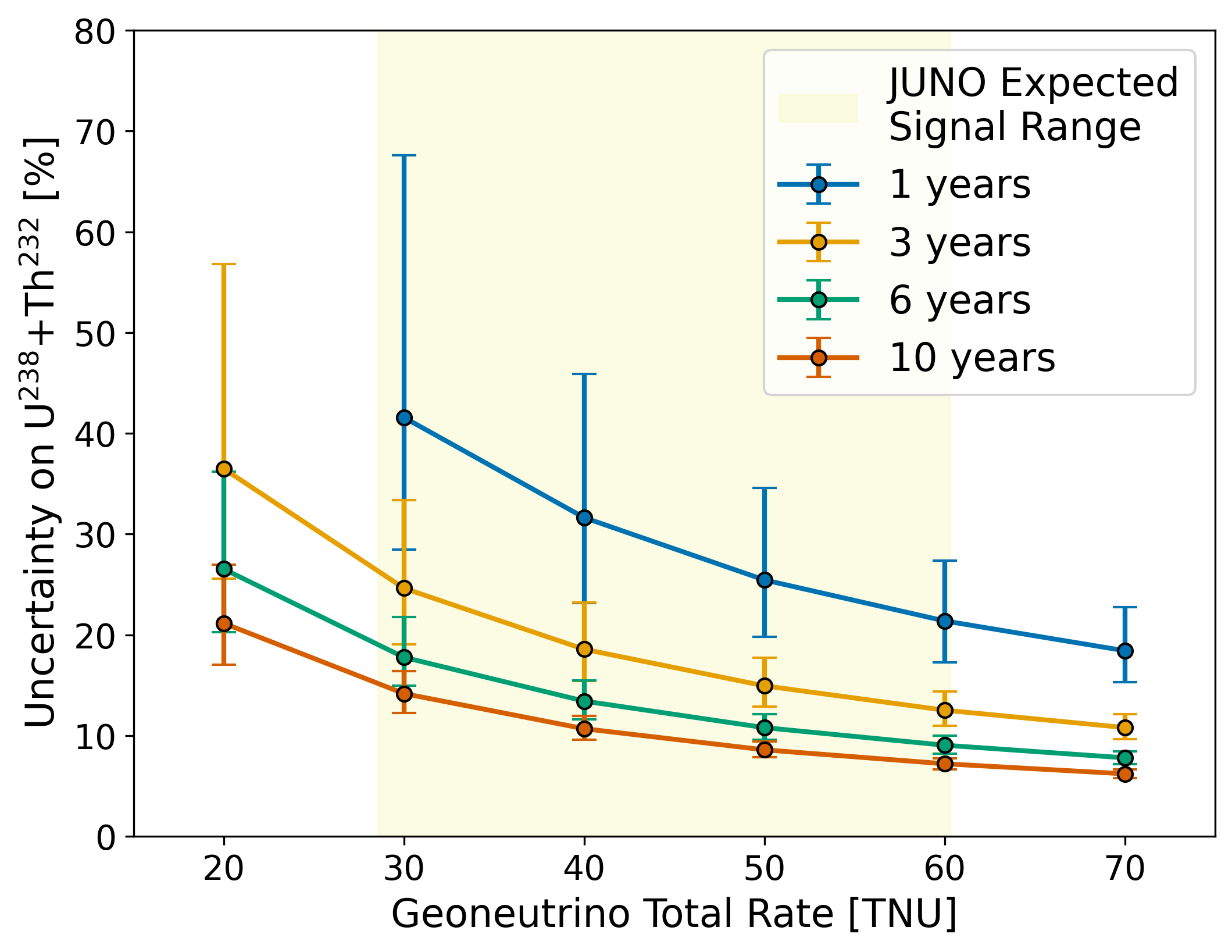}\label{fig:uthgeo_scan}}%
    \subfloat[Th/U Ratio]{\includegraphics[width=0.45\textwidth]{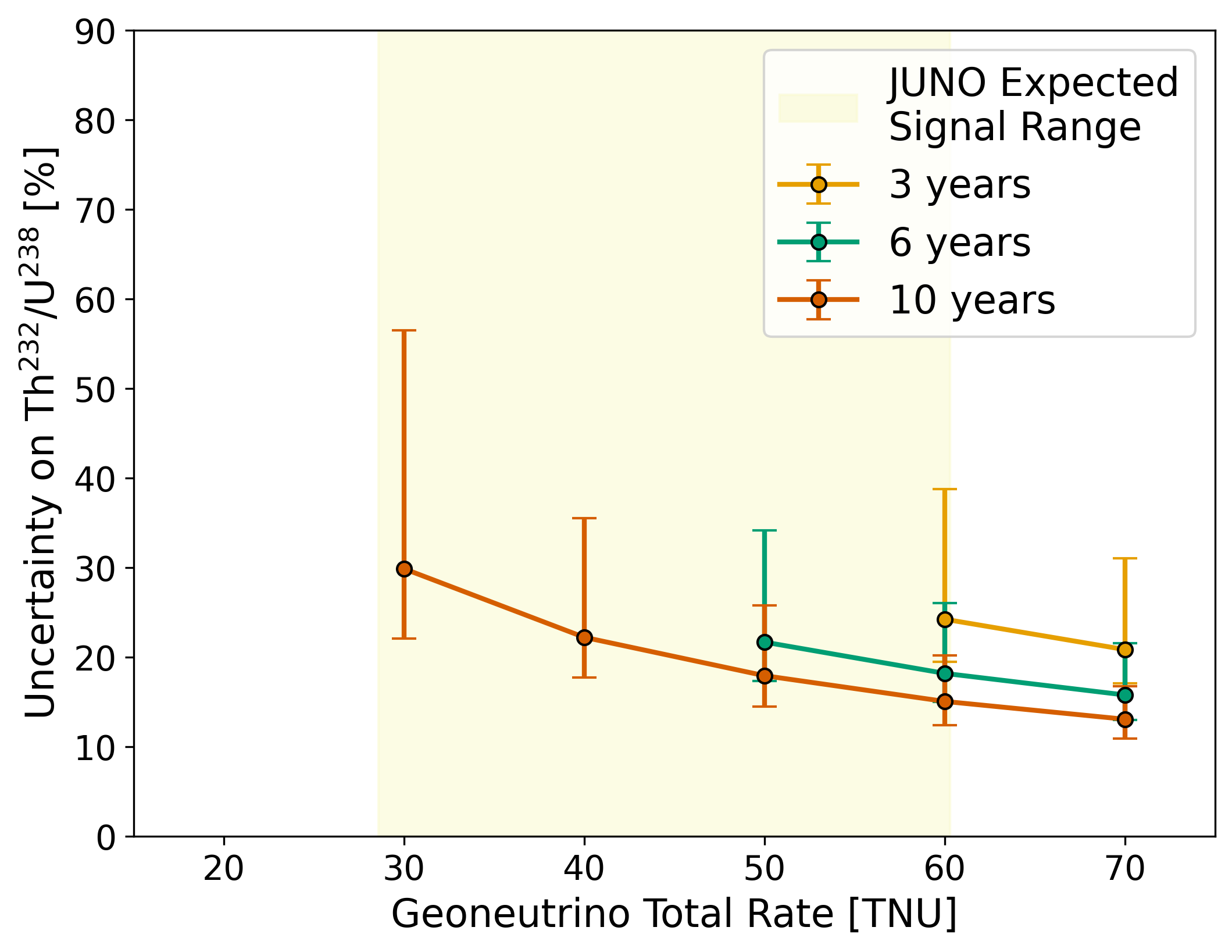}\label{fig:uthgeo_scan}}%
    \caption{Expected JUNO sensitivity to geoneutrino signal from Uranium (a), Thorium (b), their combination (c), and the Th/U ratio (d) as a function of the total geoneutrino signal. The solid blue, orange, green, and red curves represent exposures of 1, 3, 6, and 10 years, respectively. The vertical yellow shaded area indicates the 1$\sigma$ range from Table~\ref{tab:signals_tot}, encompassing the spread of different BSE models as well as various predictions for the lithospheric contribution.}%
    \label{fig:UTh_precision}%
\end{figure}

In this section, we present JUNO's projected sensitivity to the individual geoneutrino contributions from uranium (\(^{238}\)U) and thorium (\(^{232}\)Th), as well as their sum and ratio. Unlike the previous case in Section~7.1, where a fixed Th/U signal ratio was assumed, here we allow for independent variation of the uranium and thorium components in the fit, thereby enabling a more detailed geochemical interpretation of the geoneutrino signal. 

Figure~\ref{fig:UTh_precision} shows the relative uncertainties for the uranium~(a), thorium~(b), combined~(c), and the Th/U ratio~(d), as functions of the total geoneutrino signal. Four different exposure scenarios—1, 3, 6, and 10 years—are considered. The construction of Fig.~\ref{fig:UTh_precision} follows the same methodology as Fig.~\ref{fig:totageo_scan}. Points corresponding to fits with geoneutrino signals consistent with zero at less than 3\(\sigma\) significance were excluded, the central values and error bars represent the medians and ±34\% quantiles of the relative uncertainty distributions, respectively. As we can see, distinguishing the uranium and thorium contributions requires large statistics, as their spectral differences mainly appear in the high-energy tail of the geoneutrino spectrum. The uranium spectrum extends up to 3.27 MeV and dominates the region above 2.25 MeV, whereas the thorium spectrum ends at 2.25 MeV. Consequently, the uranium component can be identified earlier, with a statistically significant measurement feasible after about 3--4 years of data taking. In contrast, the thorium signal is harder to extract, as its spectrum is largely degenerate with the low-energy portion of the uranium spectrum and contributes only about 20–30\% to the total geoneutrino rate. For a total geoneutrino signal of 40~TNU, JUNO is expected to achieve a precision of $23.3_{-4.9}^{+7.0}$\% on the uranium component and  $37.0_{-10.6}^{+23.1}$\% on thorium after 6 years of exposure. After 10 years, these uncertainties further improve to 18.7$_{-2.9}^{+4.2}$\%  and $30.60_{-7.40}^{+14.41}$\%, respectively.

JUNO's exceptional event statistics thus offer a unique opportunity to disentangle uranium and thorium geoneutrino components with high statistical significance and to constrain the Th/U signal ratio. As shown in panel~(d) of Fig.~\ref{fig:UTh_precision}, this measurement can take up to 10 years,  depending on the total geoneutrino signal.
At 10 years and a 40~TNU signal, JUNO could measure the Th/U ratio with a precision $22.23_{-4.51}^{+13.32}$\%, providing valuable constraints on the global distribution of HPEs and Earth formation processes.

\begin{figure}[t]%
    \centering
{\includegraphics[width=0.5\textwidth]{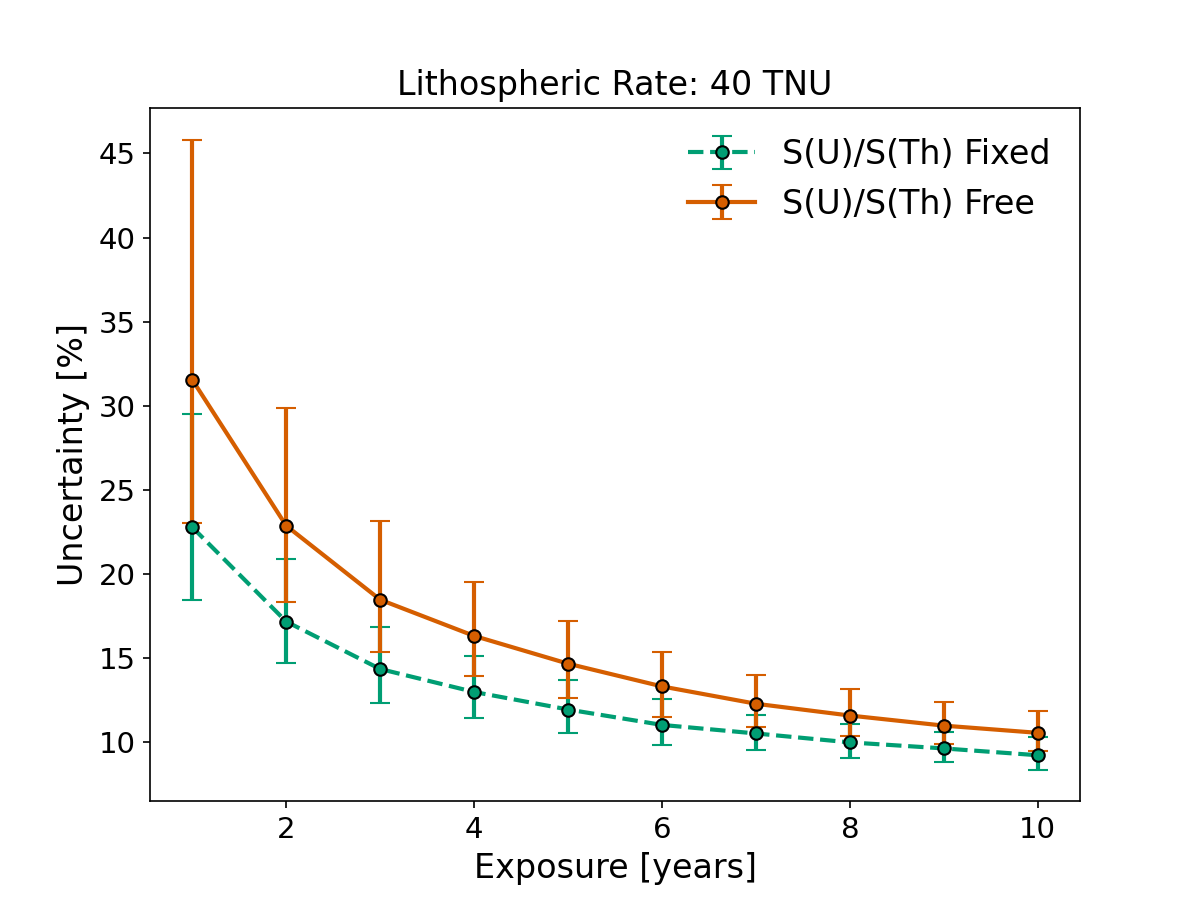}}
    \caption {Direct comparison of the expected JUNO sensitivity to the total geoneutrino signal (40\,TNU) as a function of exposure, for two fitting approaches: one with the Th/U ratio fixed to the chondritic value (green dashed line), and the other with the Th/U ratio treated as a free parameter (orange solid line).}%
    \label{fig:exposure_geo}%
\end{figure}

The combined uncertainty on the sum of U and Th geoneutrinos remains lower than the individual components due to the partial compensation of anti-correlated uncertainties. This is demonstrated for total geoneutrino signal of 40\,TNU at Fig.~\ref{fig:exposure_geo}. For example, at 6 years, the total uncertainty on $S$(U+Th) is $13.4_{-1.8}^{+2.0}$\%. For comparison, assuming a fixed Th/U ratio reduces the uncertainty to $11.0_{-1.2}^{+1.5}$\% (see Section~\ref{sec:ResultsUThFixed}). While the impact of fitting U and Th separately on the total signal precision is modest for longer exposures and higher signal levels, it becomes more relevant at lower exposures or in scenarios with reduced geoneutrino rates. For example, after just 1 year with the same 40~TNU signal, the uncertainty on \( S(\mathrm{U+Th}) \) increases to $31.6_{-8.5}^{+14.3}$ when U and Th are left free to vary, compared to $22.7_{-4.3}^{+6.7}$ for a fixed Th/U fit.

These results highlight JUNO’s potential to go beyond a total geoneutrino flux measurement and provide isotopic resolution critical for Earth science applications and directly test the Earth's Th/U ratio against model predictions.

\subsection{Mantle Signal}
\label{sec:ResultsMantle}

\begin{figure}[t]%
    \centering
    \subfloat[]{\includegraphics[width=0.5\textwidth]{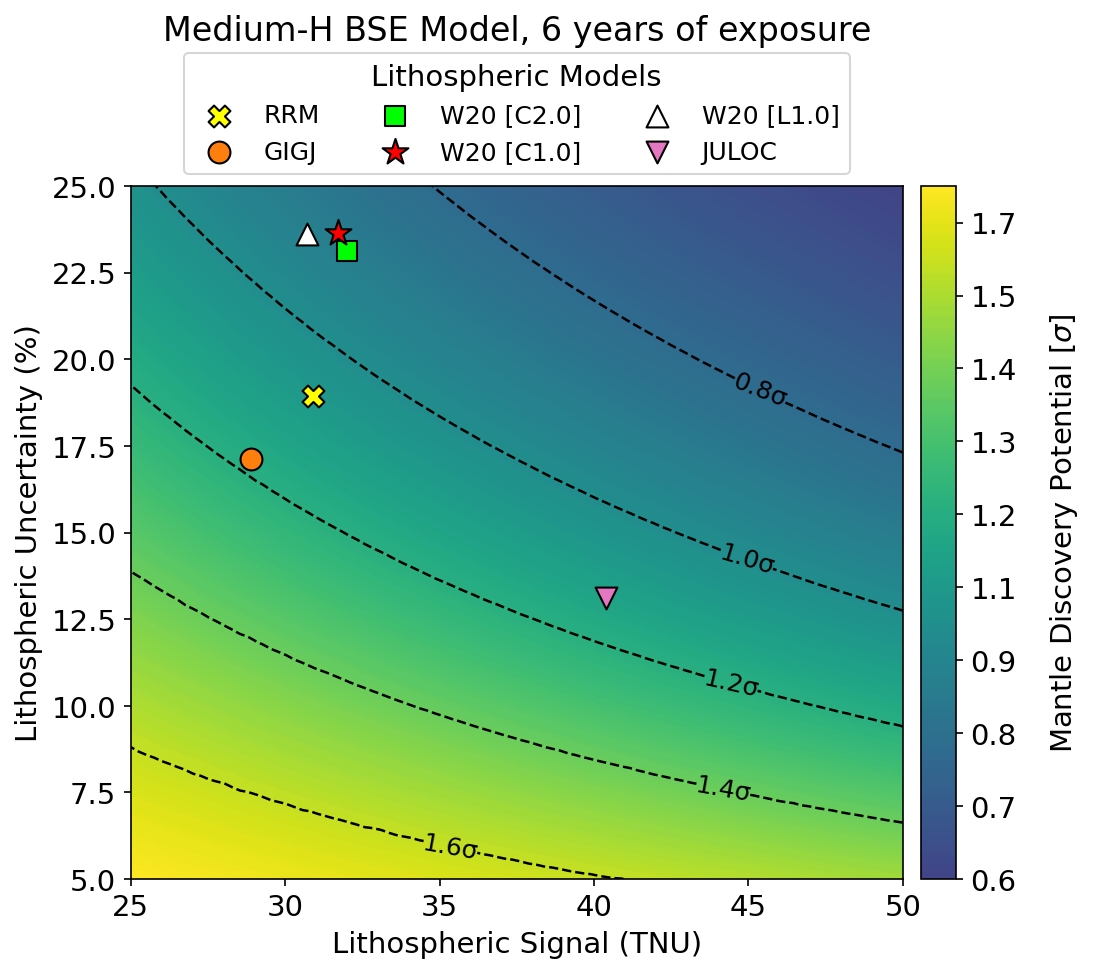}\label{fig:mantle_medH}}%
    \subfloat[]{\includegraphics[width=0.5\textwidth]{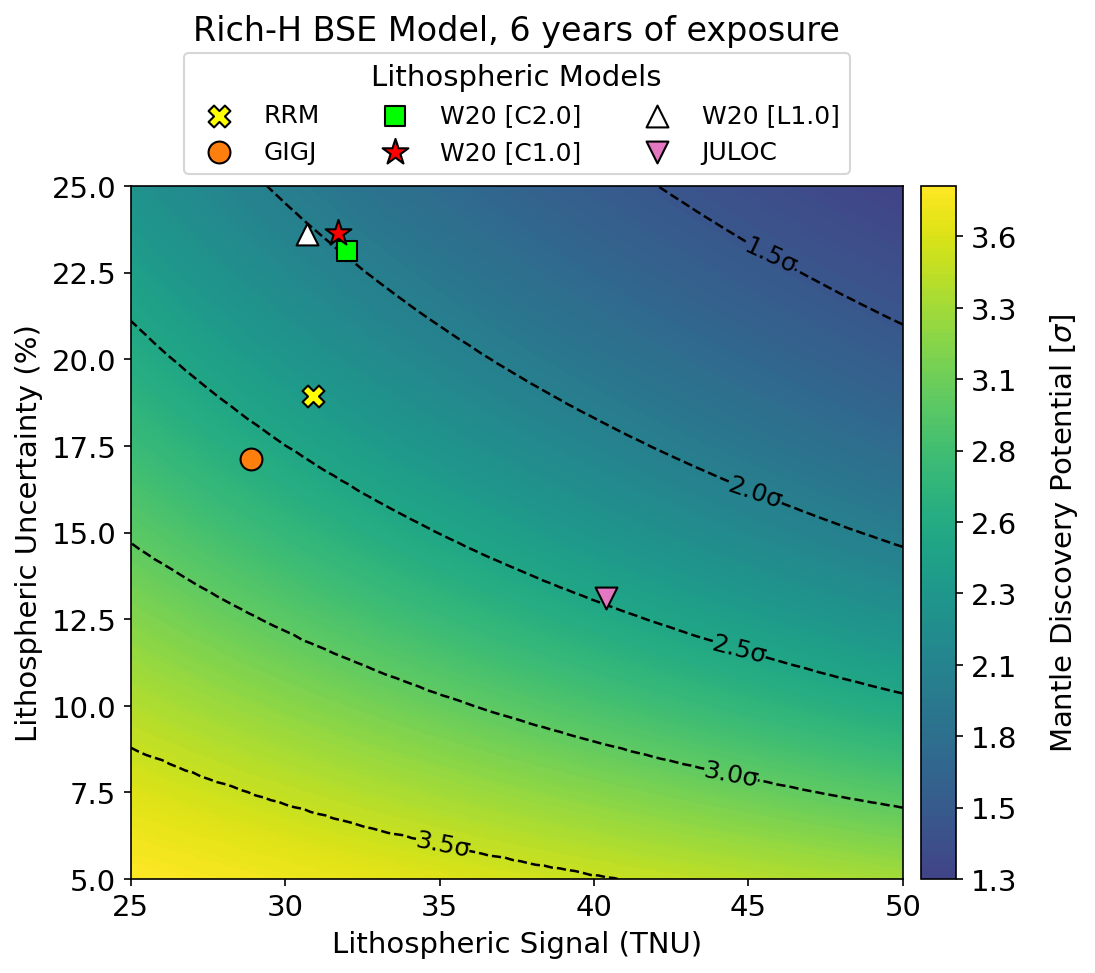}
    \label{fig:mantle_highH}}%
    \caption {Sensitivity to mantle geoneutrinos at JUNO with 6 years exposure for Medium-H (a) and Rich-H BSE (b)  models. The color scale shows the statistical significance of mantle detection as a function of the assumed lithospheric signal strength (X-axis) and its uncertainty (Y-axis). Markers indicate benchmark lithospheric models as shown in Fig.~\ref{fig:lithos_model}. }%
    \label{fig:mantle_disc_6y}%
\end{figure}

In this section we describe JUNO’s discovery potential to identify the mantle geoneutrino contribution as an excess above the lithospheric prediction. 
Since the lithosphere provides the dominant share of the total signal and carries substantial uncertainties, the sensitivity to the mantle component depends strongly on both the assumed lithospheric signal strength and the precision with which it can be constrained. 

Figure~\ref{fig:mantle_disc_6y} 
illustrates JUNO’s discovery 
potential for mantle geoneutrinos 
with 6 years of exposure, using the statistical method described in Section~\ref{subsec:mantle_strategy}. The plot shows the dependence on the assumed lithospheric signal, varied between 25 and 50\,TNU, and the precision of its constraint in the fit, varied between 5\% and 25\%. The lithospheric models from Fig.~\ref{fig:mantle_disc_6y} are indicated with different markers to benchmark our scan. Three mantle radiogenic scenarios corresponding to different BSE models (Section~\ref{subsec:mantle}) are considered. 

For the Low-H BSE model, JUNO would not be able to identify the mantle signal, so this case is omitted from Fig.~\ref{fig:mantle_disc_6y}. 
For Medium-H BSE models, JUNO’s sensitivity ranges between 1 and 1.6$\sigma$, depending on the lithospheric scenario assumed. 
A statistically significant detection of the mantle component becomes possible only for Rich-H BSE models with several years of data taking. In this scenario, a $\sim$3$\sigma$ observation could be achieved after six years of exposure, assuming a lithospheric signal of 35\,TNU constrained at the 10–15\% level. For the range of lithospheric models discussed above, the mantle signal discovery potential with 6 years of data lies between 2 and 2.5$\sigma$. 

These results highlight the importance of improving crustal models around JUNO in order to reduce the dominant lithospheric uncertainty. Such progress, combined with JUNO’s high statistics, will provide the opportunity to isolate the mantle geoneutrino flux at a continental margin site and to directly test Bulk Silicate Earth predictions.

\section{Conclusions and Outlook}
\label{sec:conclusions}

In this paper, we presented an updated evaluation of JUNO’s sensitivity to geoneutrinos, building on an improved understanding of detector performance, updated background estimates, refined reactor neutrino flux predictions, and an enhanced analysis framework. Compared to previous estimates in Refs.~\cite{An2016JUNO,Han2016}, this work includes a detailed treatment of uncertainties from both reactor and non-reactor backgrounds, allows oscillation parameters to float in the fit, and considers multiple fitting strategies. Sensitivity studies were performed using both Asimov datasets and large ensembles of toy Monte Carlo simulations.

Our sensitivity results span a broad range of signal strengths, enabling comparison with a wide variety of predictions from different geological models. From available models, the lithospheric contribution is estimated as $S_{\mathrm{LS}}$(U + Th) = $32.3^{+8.6}_{-6.6}$\,TNU. Within the Bulk Silicate Earth (BSE) framework, Poor-H, Medium-H, and Rich-H models predict total geoneutrino signals at JUNO of $36.0^{+9.6}_{-7.4}$, $41.0^{+9.5}_{-7.7}$, and $50.2^{+10.1}_{-8.5}$\,TNU, respectively, defining the expected range of geoneutrino observations. JUNO’s large 20 kton liquid scintillator target enables the detection of 0.92 to 1.95 geoneutrino events per day, surpassing the total statistics collected by all previous experiments (Borexino, KamLAND, SNO+) within a single year.

Despite the significant reactor neutrino background from the nearby ($< 300$ km) nuclear power plants producing 43.2 IBD events per day, JUNO will substantially advance our knowledge of the geoneutrino flux. For a benchmark signal of 40\,TNU, JUNO can measure the total geoneutrino signal with a precision of about $11\%$ after six years, assuming a fixed Th/U signal ratio of 0.29. If uranium and thorium are treated independently, the precision reaches around 14\%. Importantly, JUNO’s sensitivity allows for the first experimental disentanglement of uranium and thorium components: in the Medium-H scenario, the Th/U ratio can be measured with a precision of about 22\% after ten years, while in the Rich-H scenario independent determinations of uranium and thorium become feasible after roughly three years. These measurements will provide key insights into the Earth’s differentiation and thermal evolution.

Detecting the mantle contribution remains challenging due to the dominant and uncertain lithospheric signal. The mantle discovery potential depends strongly on the mantle signal strength and the adopted BSE scenario: for Medium-H models, the statistical significance after six years is limited to $\sim 1$--$1.6\sigma$, whereas for Rich-H models JUNO could achieve a $3\sigma$ detection if the crustal contribution is constrained to within 15\%. Given that the local crust around JUNO accounts for more than half of the lithospheric signal, improved geological and geophysical modeling in this region is essential to reduce uncertainties and enhance sensitivity to the mantle contribution.

In summary, JUNO’s dataset will substantially refine the measurement of the geoneutrino flux and its composition, placing the strongest constraints to date on the Earth's radiogenic heat production and Bulk Silicate Earth models. An accurate characterization of the local crust will be important to maximize the interpretive power of the results. JUNO’s data will establish a benchmark for future geoneutrino measurements and global combined analyses.

\section*{Acknowledgment}
\label{sec:ack}

We are grateful for the ongoing cooperation from the China General Nuclear Power Group. This work was supported in part by 
the Chinese Academy of Sciences, 
the National Key R\&D Program of China, 
the Guangdong provincial government,
and the Tsung-Dao Lee Institute of Shanghai Jiao Tong University in China, 
the Institut National de Physique Nucl\'{e}aire et de Physique de Particules (IN2P3) in France, 
the Istituto Nazionale di Fisica Nucleare (INFN) in Italy, 
the Fond de la Recherche Scientifique (F.R.S\-FNRS) and the Institut Interuniversitaire des Sciences Nucl\'{e}aires (IISN) in Belgium, 
the Conselho Nacional de Desenvolvimento Cient\'{\i}fico e Tecnol\'{o}gico in Brazil, the Agencia Nacional de Investigacion y Desarrollo and ANID - Millennium Science Initiative Program - ICN2019\_044 in Chile,
The European Structural and Investment Funds, the Czech Ministry of Education, Youth and Sports and the Charles University Research Centerin Czech Republic,
Deutsche Forschungsgemeinschaft (DFG), the Helmholtz Association, and the Cluster of Excellence PRISMA+ in Germany, 
the Joint Institute of Nuclear Research (JINR) and Lomonosov Moscow State University in Russia, 
the Slovak Research and Development Agency in Slovakia,
MOST and MOE in Taipei, 
the Program Management Unit for Human Resources\& Institutional Development, Research and Innovation, Chulalongkorn University, and Suranaree University of Technology in Thailand, 
the Science and Technology Facilities Council (STFC) in the UK, 
University of California at Irvine and the National Science Foundation in the US.

\pagebreak

\section{Appendix: Geoneutrino Signal Prediction at JUNO}
\label{sec:appendix}

\subsection*{Lithospheric contribution}

The lithosphere is the rigid outer shell of the Earth extends from approximately 5–10\,km in oceanic regions to 100–200\,km in continental areas ~\cite{RRM}. The lithosphere includes (i) the Earth's crust, which comprises the thin oceanic crust ($<$10\,km) and the thick continental crust (20–70\,km), and (ii) the Continental Lithospheric Mantle (CLM), which is the portion of the mantle underlying the continental crust located between the Moho and the Lithosphere-Asthenosphere boundary (Figure~\ref{fig:lithos_model}). The abundances of HPEs in the lithosphere vary significantly, with the continental crust exhibiting values on the order of 1 µg/g for U, while the oceanic crust and CLM has lower concentrations on the order of 0.1\,µg/g and 0.01\,µg/g, respectively~\cite{RRM}.

For detectors situated on the continental crust, such as JUNO, the geoneutrino signal is predominantly influenced by the lithosphere, with 40-80\% of the flux originating from this reservoir, emphasizing its crucial role in geoneutrino studies ~\cite{RRM}. This dominance arises primarily from the Local Crust (LOC), the region within a few hundred kilometers of the detector, which constitutes the largest contribution due to the geoneutrino flux's dependence on the inverse square of the distance. To enhance modeling accuracy, the LOC is analyzed separately from the Rest Of the Crust (ROC) (Figure~\ref{fig:lithos_model}), allowing for a more detailed and precise characterization of the lithospheric geoneutrino signal~\cite{Huang14,Strati17,JULOC,Sammon22}.

For this study, the geoneutrino signals from the LOC, ROC, and CLM reported in previous works were analyzed. All models employ different methodologies and approaches with variations in the size and coordinates of the LOC but remain strongly correlated due to partially shared geophysical and geochemical inputs. Below, the key aspects of each model are presented, emphasizing the key aspects for the geoneutrino signal at JUNO.

The geoneutrino signals expected at JUNO and reported in~\cite{Strati15} (Figure~\ref{tab:models}), are calculated using the global Refined Reference Earth Model \cite{RRM}, hereafter RRM, which details HPEs' abundances and distribution together with their uncertainties. Crustal thickness and uncertainty are derived from three global models (CRUST 2.0 \cite{CRUST2}, CUB 2.0 \cite{CUB}, and GEMMA \cite{GEMMA}) with crustal layer proportions, density, and elastic properties from CRUST 2.0 and sediment layer data from \cite{Laske97}. While the HPEs abundances in the sediments, oceanic crust, upper crust and CLM are assumed to be relatively homogenous based on compiled databases, the U and Th content in the deep continental crust is determined on the basis of felsic-to-mafic ratio, inferred from seismic velocities. The LOC is defined as a 6\textdegree\ $\times$ 4\textdegree\ region (20\textdegree--24\textdegree\ N, 109\textdegree--115\textdegree\ E) nearly centered on JUNO.

GIGJ (GOCE Inversion for Geoneutrinos at JUNO) \cite{GIGJ} is a 3D numerical model (cellsize: 50$\times$50$\times$0.1 km) derived by inverting GOCE gravimetric data over a 6\textdegree\ $\times$ 4\textdegree\ region exactly centered at JUNO. The a priori model integrates deep seismic sounding profiles, receiver functions, teleseismic P-wave velocity models, and Moho depth maps, each utilized according to their accuracy and spatial resolution. The unitary geoneutrino signals calculated using the site-specific subdivision of the crust of GIGJ (Table 2 of \cite{GIGJ}) combined with the HPE abundances from RRM, provided the geoneutrino signal of the LOC. To determine the total lithospheric signal, the contributions from the ROC and the CLM were added, both computed based on the RRM (Table~\ref{tab:models}). 

Wipperfurth et al. \cite{W20} present reference models for lithospheric geoneutrino signals using geophysical information from various sources, including CRUST2.0 \cite{CRUST2}, CRUST1.0 \cite{CRUST1}, and LITHO1.0 \cite{LITHO}. The contributions of the LOC (coincident with the extent defined by \cite{Strati15}), ROC, and CLM to the geoneutrino signal at JUNO are provided considering the three different geophysical models. The HPE abundances are derived using a strategy similar to the RRM, based on probability density functions of compressional wave velocity and SiO$_2$ abundance for the deep continental crust and on average global abundances for the remaining lithospheric reservoirs.

JULOC \cite{JULOC} is a 3D model of the LOC represented by the closest 10\textdegree\ $\times$ 10\textdegree\ crust surrounding the detector (17\textdegree--27\textdegree\ N, 107.5\textdegree--117.5\textdegree\ E). The geophysical inputs include a 3-D shear wave velocity model inverted from seismic ambient noise tomography. The crust is divided into upper, middle, and lower layers, with the upper crust further divided into a top layer of sedimentary rocks and granite intrusions, and a bottom layer of Precambrian basement rocks. The geochemical inputs are based on U and Th abundances from over 3000 rock samples in the region and are statistically evaluated for each crustal layer, with the uppermost layer having higher U and Th abundances. Besides the geoneutrino signal for the LOC, \cite{JULOC} reported also the signals produced by the ROC (modeled with CRUST 2.0 and compiled geochemical data) and by the CLM, inherited by~\cite{Strati15} (Table~\ref{tab:models}).

\begin{table}[t]
    \centering
    \renewcommand{\arraystretch}{1.3}
   \begin{tabular}{lccccc}
        \hline\hline
        \textbf{Model} & \textbf{LOC Area extension} & \multicolumn{3}{c}{\textbf{S(U + Th) [TNU]}} & \textbf{Reference} \\
        & (Long x Lat) & LOC & ROC & CLM & \\
        \hline\hline
        RRM & $6^\circ \times 4^\circ$ & $16.4^{+3.0}_{-2.6}$ & $11.7^{+2.3}_{-1.9}$ & $2.1^{+2.9}_{-1.3}$ & ~\cite{Strati15}\\
        GIGJ & $6^\circ \times 4^\circ$ & $14.4^{+2.5}_{-2.4}$ & $11.7^{+2.2}_{-1.9}$ & $2.1^{+2.9}_{-1.3}$ &~\cite{GIGJ} \\
        W20 [Crust\_2.0] & $6^\circ \times 4^\circ$ & $16.9^{+4.3}_{-3.4}$ & $11.8^{+3.3}_{-2.6}$ & $1.7^{+2.9}_{-1.1}$ & ~\cite{W20}\\
        W20 [Crust\_1.0] & $6^\circ \times 4^\circ$ & $16.7^{+4.4}_{-3.5}$ & $11.7^{+3.3}_{-2.6}$ & $1.7^{+2.9}_{-1.1}$& ~\cite{W20}\\
        W20 [Lithos1.0] & $6^\circ \times 4^\circ$ & $17.1^{+4.5}_{-3.6}$ & $12.7^{+3.6}_{-2.8}$ & $0.6^{+0.9}_{-0.4}$ & ~\cite{W20}\\
        JULOC & $10^\circ \times 10^\circ$ & $28.5^{+4.5}_{-4.5}$ & $9.8^{+1.7}_{-1.7}$ & $2.1^{+2.9}_{-1.3}$ &~\cite{JULOC}\\
        \hline\hline
    \end{tabular}
    \caption{Summary of geoneutrino signal (S (U+Th)) contributions from the Local Crust (LOC), Rest of the Crust (ROC), and Continental Lithospheric Mantle (CLM) for the different models. The LOC area extension (longitude × latitude) is reported for each model.}
    \label{tab:models}
\end{table}

The geoneutrino signal predicted by the JULOC model is significantly higher than those of the other models (Table~\ref{tab:signals_ls}). While most models report a lithospheric geoneutrino signal $S_\text{LS}(\text{U}+\text{Th})$ with a central value in the range of 28.9~TNU to 32.0~TNU, with an average relative uncertainty of about 20\%, JULOC stands out with a much larger signal of $S_\text{LS}(\text{U}+\text{Th}) = 40.4~\text{TNU}$ and a lower relative uncertainty ($\sim 10\%$) (Figure~\ref{fig:sig_lithos}). The ratios of S(Th)/S(U) remain consistent across all models, with values clustering around 0.30–0.31. This stability reflects the strong correlation between U and Th contributions and highlights the uniformity in the assumed Th/U abundance ratio across the models.

\begin{table}[t]
    \centering
    \renewcommand{\arraystretch}{1.3}
    \begin{tabular}{ccccc}
       \hline\hline
       Model & S$_{LS}$(U) [TNU] & S$_{LS}$(Th) [TNU] & S$_{LS}$(U+ Th) [TNU] & S$_{LS}$(Th)/S$_{LS}$(U)\\
       \hline\hline
        RRM & $23.2^{+5.9}_{-4.8}$ & $7.3^{+2.4}_{-1.5}$ & $30.9^{+6.5}_{-5.2}$ & 0.31\\
        GIGJ & $22.0^{+5.4}_{-4.1}$ & $6.6^{+1.7}_{-1.2}$ & $28.9^{+5.5}_{-4.4}$ & 0.30\\ W20 [Crust 2.0] & $24.5^{+6.4}_{-5.1}$ & $7.4^{+2.1}_{-1.7}$ & $32.0^{+8.2}_{-6.6}$ & 0.30 \\
        W20 [Crust 1.0] & $24.2^{+6.5}_{-5.1}$ & $7.4^{+2.2}_{-1.7}$ & $31.7^{+8.4}_{-6.6}$ & 0.31 \\
        W20 [Litho 1.0] & $23.6^{+6.4}_{-5.0}$ & $7.1^{+2.0}_{-1.6}$ & $30.7^{+8.1}_{-6.4}$ & 0.30\\
        JULOC & $30.3^{+4.9}_{-4.4}$ & $9.2^{+1.7}_{-1.4}$ & $40.4^{+5.6}_{-5.0}$ & 0.30\\
        Combined & $24.7^{+6.7}_{-5.5}$ & $7.5^{+2.2}_{-1.8}$ & $32.3^{+8.6}_{-6.6}$ & 0.30\\
       \hline\hline
    \end{tabular}
    \caption{Predicted geoneutrino signals (S(U+Th)) at JUNO and its components (S(U)) and (S(Th)) for the different lithospheric models together with the corresponding ratio S(Th)/S(U). The combined signal represents a statistically integrated result, incorporating contributions from all models. Uncertainties are reported as $1\sigma$ interval}
    \label{tab:signals_ls}
\end{table}

Note that each of the considered works, published between 2015 and 2020, adopt different oscillation parameters based on the available values at the time of the publication. Although subsequent updates to these parameters may introduce slight variations in the geoneutrino signal calculation, their impact on the average oscillation probability is {small} compared to the uncertainties arising from geological inputs and assumptions.

Since the six models used to estimate the geoneutrino signal share common features (e.g., global geophysical models, U and Th abundances in crustal layers), it is necessary to account for strong correlations when calculating a single central value and the corresponding uncertainty for the expected lithospheric signal at JUNO. Specifically, $10^6$ values were generated for each geoneutrino model based on its probability density function, representing the signal distributions predicted by the respective models. The generated samples were pooled into an aggregated dataset, from which the median and the $16^{th}$ and $84^{th}$ percentiles (defining the $1\sigma$ interval) were extracted.

\subsection*{Mantle contribution}

Once the total U and Th content is fixed for a given BSE class, and the lithospheric U and Th abundances are determined ($H_\text{LS} = 6.9_{-1.2}^{+1.6}~\text{TW}$),~\cite{RRM}), the residual amounts of U and Th in the mantle are calculated in by subtracting the lithospheric contribution from the total BSE estimates (Table 29 of~\cite{Bell21}). The mantle radiogenic heat obtained for each class (Table~\ref{tab:mantle_signal_text}) reflects the assumptions regarding the Earth’s primordial composition and the partitioning of HPEs.

\begin{figure}[t]
    \centering
    \includegraphics[width=0.6\linewidth]{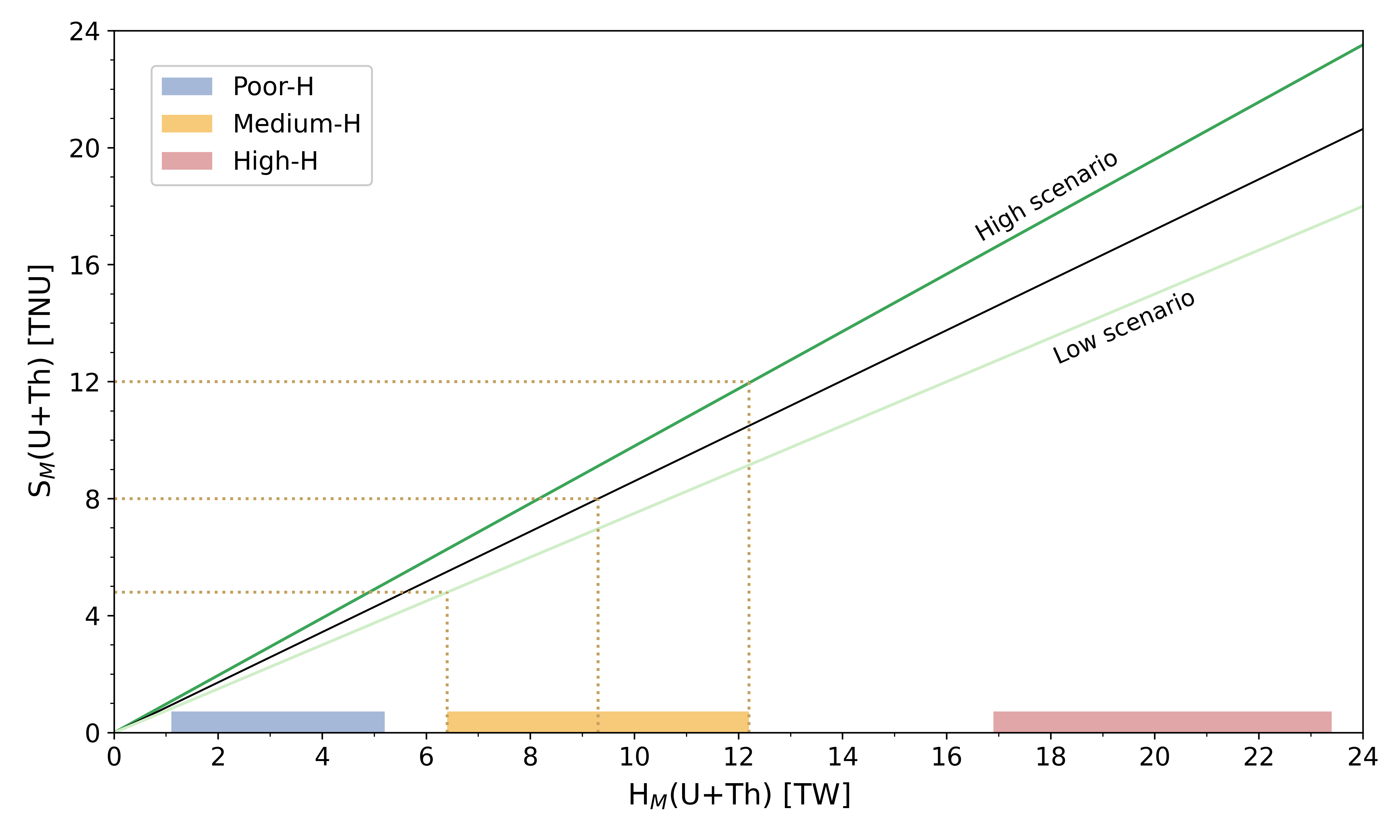}
    \caption{Mantle geoneutrino signal ($S_\text{M}(\text{U} + \text{Th})$) as a function of U and Th mantle radiogenic heat $H_\text{M}(\text{U} + \text{Th})$. The blue, yellow, and red bands on the X-axis correspond to the 68\% coverage interval of the mantle radiogenic heat ($H_\text{M}(\text{U} + \text{Th})$) of Poor-H, Medium-H, and Rich-H models (Table~\ref{tab:mantle_signal_text}), respectively. The area between the green lines denotes the full range allowed between a homogeneous mantle (high scenario, $\beta = 0.98~\text{TNU}~\text{TW}^{-1}$) and a unique enriched layer just above the core-mantle boundary (low scenario, $\beta = 0.75~\text{TNU}~\text{TW}^{-1}$); the inclined black line identifies the average coefficient ($\beta = 0.86~\text{TNU}~\text{TW}^{-1}$). The dashed yellow line depicts, as an example, the method adopted to derive the central value of the $S_\text{M}(\text{U} + \text{Th})$ together with the upper and lower bounds for the Medium-H class.}
    \label{fig:sig_mantle}
\end{figure}

The expected mantle geoneutrino signal for each BSE class is determined using the linear relation $S_\text{M}(\text{U}+\text{Th}) = \beta \cdot H_\text{M}(\text{U}+\text{Th})$, where $\beta$ represents the proportionality coefficient that depends solely on the spatial distribution of uranium (U) and thorium (Th) in the mantle. The value of $\beta$ ranges from $\beta_\text{low} = 0.75~\text{TNU/TW}$ when HPEs are concentrated in a thin layer above the core-mantle boundary to $\beta_\text{high} = 0.98~\text{TNU/TW}$ for a homogeneous mantle distribution; the average value is $\beta = 0.86~\text{TNU/TW}$ ~\cite{BOREX20, Bell21}. Using the radiogenic heat values $H_\text{M}(\text{U}+\text{Th})$ for the three BSE classes, the corresponding geoneutrino signals $S_\text{M}(\text{U}+\text{Th})$ can be plotted (Figure~\ref{fig:sig_mantle}). Specifically, the central value of the geoneutrino signal is determined by using the median $H_\text{M}(\text{U}+\text{Th})$ and applying the average $\beta$. The upper limit of uncertainty for $S_\text{M}(\text{U}+\text{Th})$ is obtained by increasing $H_\text{M}(\text{U}+\text{Th})$ by one standard deviation and applying $\beta_\text{high}$. Conversely, the lower uncertainty limit is calculated by decreasing $H_\text{M}(\text{U}+\text{Th})$ by one standard deviation and using $\beta_\text{low}$ (Figure~\ref{fig:sig_mantle}). This method provides a comprehensive approach to estimate the expected geoneutrino signal ranges for different mantle compositions and distribution scenarios.

The individual signal contributions of uranium $S_\text{M}(\text{U})$ and thorium $S_\text{M}(\text{Th})$ to the mantle geoneutrino signal for each BSE class are computed using the relation $S_\text{M}(\text{Th})/S_\text{M}(\text{U}) = 0.069 \times \left[M_\text{Th}/M_\text{U}\right]$~\cite{BOREX20}, where $M_\text{Th}$ and $M_\text{U}$ are the thorium and uranium mass in the mantle. By applying the mass ratios from Table 29 of~\cite{Bell21} and the mantle geoneutrino signal $S_\text{M}(\text{U}+\text{Th})$, the individual signals $S_\text{M}(\text{U})$ and $S_\text{M}(\text{Th})$ can be derived together with their ratio $S_\text{M}(\text{Th}) / S_\text{M}(\text{U})$, as it was shown in  Table~\ref{tab:mantle_signal_text}. Since the main uncertainties in the mantle signal arise from variations in the total U and Th content across different BSE classes and their spatial distribution within the mantle, the uncertainty in the $M_\text{Th}/M_\text{U}$ ratio is negligible in the overall error budget.

\bibliographystyle{plainnat}
\bibliography{main}

\end{document}